# The Recycler Electron Cooler

*A. Shemyakin\*\* and L. R. Prost*

*FNAL\*, Batavia, IL 60510, USA*

*Abstract*
The Recycler Electron cooler was the first (and so far, the only) cooler working at a relativistic energy ($\gamma = 9.5$). It was successfully developed in 1995-2004 and was in operation at Fermilab in 2005-2011, providing cooling of antiprotons in the Recycler ring. This paper describes the cooler, difficulties in achieving the required electron beam parameters and the ways to overcome them, cooling measurements, and details of operation.

___

**Table of Contents**

# 1. Introduction

Electron cooling is the method of increasing the phase density of "hot' heavy charge particles, ions or antiprotons, through Coulomb interaction with a "cold" electron beam propagating with the same average speed. The method was proposed by G. Budker in 1967 [1], successfully tested in 1974 with low-energy protons [2], and later implemented at a dozen of storage rings (see, for example the review [3]) at non-relativistic energies $E_e$ <300 keV.

An electron cooler was envisioned as an important part of the Recycler ring upgrade already in the Recycler Technical Design Report [4]. The main cooler parameters were chosen according to the then stated Recycler goals: store antiprotons coming from the Accumulator, "recycle" antiprotons left over from Tevatron stores, and prepare bunches for Tevatron shots. Because of the longitudinal injection scheme of the Recycler, the main emphasis was made on longitudinal cooling. With a typical requirement of tens of minutes for the cooling time, the scheme without a strong magnetic field in the cooling section was shown to be satisfactory. Later changes, most notably the decision to do not reuse the antiprotons left over from the Tevatron and the decreased emittances of the bunches coming from the Accumulator, relaxed the operational requirements for the electron cooling strength, which allowed operation with a good safety factor. As soon as the issues with reliable recirculation of the electron beam were resolved, for the first time, relativistic electron cooling was demonstrated [5] and within days was put into operation.

This paper introduces formulae for non-magnetized electron cooling (Section 2); discusses the setup and reasoning behind the choice of the cooler's scheme (Section 3); and then goes through the main difficulties with the realization of the project: electron beam transport (Section 4), stability of the beam recirculation (Section 5), electron angles in the cooling section (Section 6), and energy matching (Section 7). Section 8 describes the cooling force measurements and results, and then we conclude.

For brevity, sometimes we will refer to the Recycler Electron cooler as REC.

# 2. Electron cooling formulae

A heavy charged particle moving in a free electron gas with a velocity distribution $f_e(\vec{v}_e)$ experiences a friction force that in a model of binary collisions can be written following Ref. [6]:

$$\vec{F}_b(\vec{V}_p) = -\frac{4\pi e^4 n_{eb}}{m_e}\eta \int L_C \frac{f_e(\vec{v}_e)}{(\vec{V}_p - \vec{v}_e)^2} \frac{\vec{V}_p - \vec{v}_e}{|\vec{V}_p - \vec{v}_e|} d^3 v_e, \qquad (1)$$

where $n_{eb}$ is the electron density in the beam rest frame, $m_e$ the electron mass, $e$ the elementary charge, $\vec{V}_p$ the velocity of the heavy particle, and $\eta = L_{cs}/C$ indicates the portion of the ring circumference $C$ occupied by the cooling section of length $L_{cs}$. $L_C$ is the Coulomb logarithm

$$L_C = \ln\left(\frac{\rho_{max}}{\rho_{min}}\right) \qquad (2)$$

with the minimum and maximum impact parameters, $\rho_{min}$ and $\rho_{max}$, in the Coulomb logarithm defined as

$$\rho_{min} = \frac{e^2}{m_e(\vec{V}_p - \vec{v}_e)^2}, \quad \rho_{max} = \min\left\{R_D, R_e, |\vec{V}_p - \vec{v}_e|\cdot \tau_f\right\}. \qquad (3)$$

The maximum impact parameter is determined by the electron beam radius $R_e$, (typically the case in REC), the Debye radius $R_D$, or the relative displacement of particles during the flight time through the cooling section $\tau_f = \dfrac{L_{cs}}{\gamma \beta c}$, where $\gamma$ and $\beta$ are the relativistic factors of co-propagating particles in the lab frame, whichever is the greatest.

Note that the presence of a strong longitudinal magnetic field ~ 1 kG, standard for low-energy coolers, makes cooling dynamics significantly more complicated (see e.g.: [6], [3]) and typically provides significantly stronger cooling. For the REC, we did not find a way to reliably simulate the possible contribution to cooling of this effect, where electrons are immersed into a much weaker field of $B_{CS} = 105$ G (see

Table 1) but believe that Eq. (1) is a reasonable approximation. One may speculate that the magnetic field doesn't affect collisions with "small" impact parameters $\rho$ i.e.

$$\rho \leq \rho_L \approx \frac{v_e \cdot m_e c}{e B_{CS}} \quad (4)$$

but enhances cooling at "large" impact parameters

$$\rho > \rho_L \quad (5)$$

by "suppressing" the transverse electron velocities contribution. For typical REC parameters, $\rho_{min} \approx 1 \cdot 10^{-7}$ cm, $\rho_L \approx 0.02$ cm, $\rho_{max} \approx 0.2$ cm, and the value of the Coulomb logarithm is significantly larger for the region of Eq. (4), $ln(\rho_L / \rho_{min}) \approx 12$, than $ln(\rho_{max} / \rho_L) \approx 2.3$ for large impact parameters. An enhancement of the cooling force by the magnetic field for large impact parameters is unlikely to offset this difference because the typical value of the antiproton velocity (in the beam frame) is only a factor of 2-3 lower than the electrons'. An even more convincing argument comes from measurements, where a strong dependence of cooling properties on the transverse electron angles was observed.

Assuming a Gaussian distribution for all velocity components

$$f(\vec{v}_e) = \frac{1}{(2\pi)^{3/2} \sigma_{ex} \sigma_{ey} \sigma_{ez}} \exp\left(-\frac{v_{ex}^2}{2\sigma_{ex}^2} - \frac{v_{ey}^2}{2\sigma_{ey}^2} - \frac{v_{ez}^2}{2\sigma_{ez}^2}\right) \quad (6)$$

and neglecting variations of the Coulomb logarithm in order to take it out of the integrand, Eq. (3.1) can be reduced to a single integral (so-called Binney formula, e.g.: Ref. [7]). For example, the expression for the longitudinal component of the cooling force in the beam frame becomes

$$F_{bz}(V_{px}, V_{py}, V_{pz}) = -\frac{4\Lambda}{\sqrt{\pi}} \cdot \int_0^\infty dt \frac{\exp\left(-\dfrac{t^2 V_{px}^2}{1 + 2 \cdot \sigma_{ex}^2 t^2} - \dfrac{t^2 V_{py}^2}{1 + 2 \cdot \sigma_{ey}^2 t^2} - \dfrac{t^2 V_{pz}^2}{1 + 2 \cdot \sigma_{ez}^2 t^2}\right)}{\sqrt{(1 + 2 \cdot \sigma_{ex}^2 t^2)(1 + 2 \cdot \sigma_{ey}^2 t^2)(1 + 2 \cdot \sigma_{ez}^2 t^2)}} \cdot \frac{V_{pz} t^2}{1 + 2 \cdot \sigma_{ez}^2 t^2} \quad (7)$$

$$\Lambda \equiv \frac{4\pi e^4 n_{eb}}{m_e} \eta L_C$$

This formula can be simplified further for the analysis of the longitudinal cooling force measurements (see section 7.8.8). These measurements use a pencil antiproton beam with a large average momentum offset, so that the transverse antiproton velocity components can be neglected.

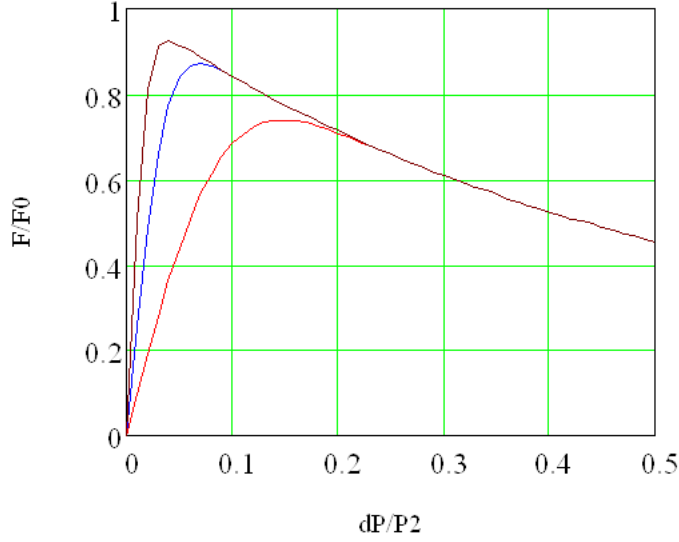

Figure 1. Longitudinal force as a function of the momentum offset calculated with Eq. (8). Vertical axis - $F_{lz}(\Delta p_p)/F_0$, horizontal axis - $\Delta p_p/p_2$. The curves show calculations for different ratio of electron velocities in the beam frame $p_1/p_2 \approx \sigma_{ez}/\sigma_{et}$: red- 10, blue- 25, brown- 50.

Assuming also $\sigma_{ex} = \sigma_{ey} \equiv \sigma_{et}$, it is convenient to use Eq.(7) to fit the longitudinal cooling force as a function of momentum offset in the lab frame in the form of [8]

$$F_{lz}(\Delta p_p) = -F_0 \frac{2}{\sqrt{\pi}} \int_0^{\frac{\Delta p_p}{p_1}} \frac{e^{-u^2} u^2}{u^2 + \left(\frac{\Delta p_p}{p_2}\right)^2} du \qquad (8)$$

with the parameters related to the lab-frame electron beam properties as follows:

$$\begin{aligned} p_1 &= \delta W_e \cdot \sqrt{2}\, \frac{M_p}{\beta m_e c} \\ p_2 &= \vartheta_t \cdot \sqrt{2}\, \gamma^2 \beta c M_p \\ F_0 &= \frac{\Lambda}{\sigma_{et}^2 - \sigma_{ez}^2} = \frac{n_{el}}{\vartheta_t^2} \cdot \frac{4\pi \cdot e^4 \eta \cdot L_c}{m_e c^2 \gamma^3 \beta^2} \end{aligned} \qquad (9)$$

where

$\vartheta_t = \frac{1}{\gamma \beta c}\sqrt{\sigma_{et}^2 - \sigma_{ez}^2}$ - an effective angle,

$\vartheta_e = \frac{\sigma_{et}}{\gamma \beta c}$ -  1D r.m.s. electron angle in the cooling section,

$\delta W_e = p_e \sigma_{ez}$ - r.m.s. scatter of the electron energy,

$n_{el} = \gamma\, n_{eb}$ - electron density in the lab frame,

$M_p$ is the proton mass, and $p_e$ is the electron momentum. These parameters used in Eq.(8) are convenient for fitting purposes because $F_0$ determines the maximum of the curve (the force maximum approaches $F_0$ at $\sigma_{et} \gg \sigma_{ez}$), the parameter $p_1$ is approximately the position of the curve's maximum, and $p_2$ characterizes the curve's drop rate after the maximum (**Error! Reference source not found.**).

Under the assumptions used for Eq.(7) and $\sigma_{ex} = \sigma_{ey} \equiv \sigma_{et}$, the cooling rates of a coasting antiproton beam with a Gaussian distribution

$$f_p(\vec{V}_p) = \frac{1}{(2\pi)^{3/2} \sigma_{px} \sigma_{py} \sigma_{pz}} \exp\left(-\frac{V_{px}^2}{2\sigma_{px}^2} - \frac{V_{py}^2}{2\sigma_{py}^2} - \frac{V_{pz}^2}{2\sigma_{pz}^2}\right) \quad (10)$$

with $\sigma_{px} = \sigma_{py}$ can be expressed through elementary functions [8]:

$$\dot{\sigma}_{pz} = -\sqrt{\frac{2}{\pi}} \frac{\Lambda \cdot \sigma_{pz}}{M_p \sqrt{\sigma_{ez}^2 + \sigma_{pz}^2} (\sigma_{ex}^2 + \sigma_{px}^2)} \cdot f_{long}(\alpha); \quad \frac{d}{dt_b}(\sigma_{px}^2) = -\sqrt{\frac{2}{\pi}} \frac{\Lambda \cdot \sigma_{px}^2}{M_p (\sigma_{ex}^2 + \sigma_{px}^2)^{3/2}} \cdot f_{tr}(\alpha);$$

$$f_{long}(\alpha) = \begin{cases} \dfrac{1 - \sqrt{\dfrac{\alpha}{1-\alpha}} \cdot \arccos(\sqrt{\alpha})}{1-\alpha}, & \alpha < 1 \\ \dfrac{1}{3}, & \alpha = 1 \\ \dfrac{\sqrt{\dfrac{\alpha}{\alpha-1}} \cdot \cosh^{-1}(\sqrt{\alpha}) - 1}{\alpha - 1}, & \alpha > 1 \end{cases} \quad ; \quad \alpha = \frac{\sigma_{ez}^2 + \sigma_{pz}^2}{\sigma_{ex}^2 + \sigma_{px}^2};$$

$$f_{tr}(\alpha) = \begin{cases} \dfrac{\arccos\sqrt{\alpha} - \sqrt{\alpha(1-\alpha)}}{2(1-\alpha)^{3/2}}, & \alpha < 1 \\ \dfrac{1}{3}, & \alpha = 1 \\ \dfrac{\sqrt{\dfrac{\alpha}{\alpha-1}} \cdot \cosh^{-1}(\sqrt{\alpha}) - 1}{\alpha - 1}, & \alpha > 1 \end{cases} \quad ; \quad (11)$$

Note that for the transverse rate, the calculated single–pass force is decreased in Eq.(11) by a factor of 2 to take into account averaging over betatron oscillations. Functions appearing in Eq.(3.9) are shown in **Error! Reference source not found.**. For small values of the parameter $\alpha \ll 1$, $f_{tr}(\alpha) \approx \pi/4 - \sqrt{\alpha}/2$, $f_{long}(\alpha) \approx 1 - \pi/2 \cdot \sqrt{\alpha}$.

For a practical implementation, Eq.(11) gives only an upper estimate for the cooling rate. The most important flaw in the model is the assumption of the electron beam density and velocity distributions being constant across the area determined by $\rho_{max}$. In the case of the Recycler cooler operational parameters, the rms radius of the antiproton beam is roughly the same as the radius of the area within the electron beam where electron cooling is effective (see section 8.2). As a result, a large portion of antiprotons travels through the cooling section outside of the electron beam hence the cooling rate decreases dramatically. With the tails of the antiproton distribution being typically fatter than Gaussian, this effect is even more pronounced. In addition, bringing the Coulomb logarithm out of the integral tends to slightly overestimate the cooling force for small relative velocities (section 8.3). Numerical comparison of the measured cooling rate with Eq.(11) is shown in Figure 30.

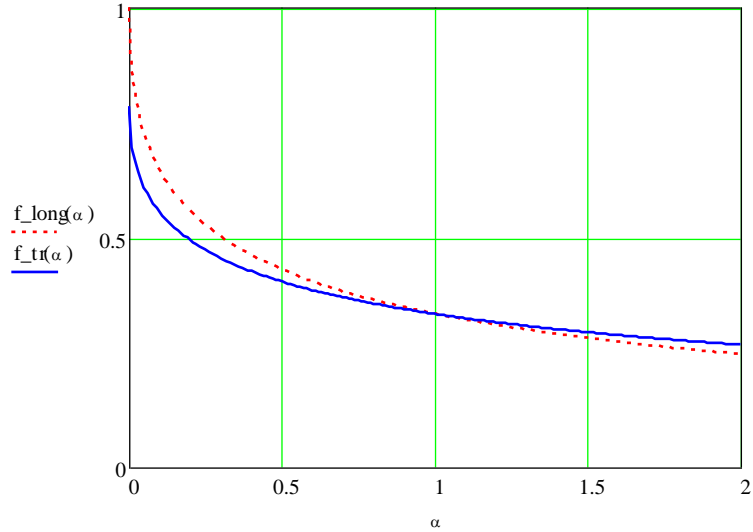

Figure 2. Functions in Eq. (11).

For a practical implementation, Eq.(11) gives only an upper estimate for the cooling rate. The most important flaw in the model is the assumption of the electron beam density and velocity distributions being constant across the area determined by $\rho_{max}$. In the case of the Recycler cooler operational parameters, the rms radius of the antiproton beam is roughly the same as the radius of the area within the electron beam where electron cooling is effective (see section 8.2). As a result, a large portion of antiprotons travels through the cooling section outside of the electron beam hence the cooling rate decreases dramatically. With the tails of the antiproton distribution being typically fatter than Gaussian, this effect is even more pronounced. In addition, bringing the Coulomb logarithm out of the integral tends to slightly overestimate the cooling force for small relative velocities (section 8.3). Numerical comparison of the measured cooling rate with Eq.(11) is shown in Figure 30.

## 3. Cooler setup

### 3.1. Electron beam design parameters and choice of the scheme

Based on preliminary cooling scenarios and estimations of the cooling rates, the design parameters were specified in Ref. [9] (reproduced in
Table 1).

Table 1 assumes a scheme with a DC electron beam, a longitudinal magnetic field at the cathode and in the cooling section, and lumped focusing in the beam transport lines. Before finally choosing this scheme, other possibilities were considered.

By the time the Recycler Electron Cooling project started, several schemes for coolers with electron energy of several MeVs or higher had been investigated ( [10], [11], [12]) (and the development of other schemes continues at present ( [13], [14])). However, none of these developments came close to demonstrating the electron beam parameters necessary for electron cooling in the Recycler.

Table 1: Electron cooling system design parameters

| Parameter | Value | Units |
|---|---|---|
| *Electrostatic Accelerator* | | |
| Terminal Voltage | 4.3 | MV |
| Electron Beam Current | 0.5 | A |
| Terminal Voltage Ripple, rms | 500 | V |
| Cathode Radius | 2.5 | mm |
| Magnetic Field at the Cathode | ≤ 600 | G |
| *Cooling Section* | | |
| Length | 20 | m |
| Solenoid Field | ≤ 150 | G |
| Vacuum Pressure | 0.1 | nTorr |
| Electron Beam Radius | 6 | mm |
| Electron Beam Divergence | ≤ 80 | μrad |

In all coolers that had been built previously, a strong (~1 kG) longitudinal magnetic field was used to transport the electron beam and enhance the cooling force, but all of them worked at non-relativistic energies $E_e$ <300 keV. Therefore, a straightforward extrapolation of the previous experience would have been a version with a higher-voltage electrostatic accelerator and a continuous strong longitudinal magnetic field from the cathode to the collector. This was seriously discussed but eventually abandoned. On one hand, the Recycler did not require the benefits from cooling enhancement obtained with a strong field. On the other hand, the beam generation scheme tested for 1 MeV [10] was not easily scalable to 4 MeV and would have required significant R&D efforts (similar to what is being presently developed for the COSY cooler at Novosibirsk [14]). Combined with more expensive beam lines and cooling section as well as higher operating costs, it was deemed undoable within a realistic budget then available and the time scale on which the project needed to be completed.

The most affordable solution in terms of time and cost seemed to be the scheme of relativistic cooling proposed for the SSC MEB [15], which assumed that focusing in the cooling section would be achieved with lumped elements. Namely, as opposed to low energy coolers, there is no longitudinal magnetic field where the beam is generated, and then short solenoidal lenses are placed periodically in the cooling section to compensate the beam divergence caused by space charge and beam emittance. This configuration allows using an industrially-manufactured electrostatic accelerator, and the cooling section is significantly cheaper than in the case with a strong magnetic field. This scheme was critically analyzed in Ref. [16]. The authors argued that the requirement of low transverse velocities in the cooling section results in a large value of the beta-function, which makes the beam susceptible to perturbations. Specifically, a drift instability due to the interaction of the beam with its wall image charges and an ion instability were pointed out as possible showstoppers. In addition to these arguments, cooling inside the lenses is ineffective because of the large azimuthal velocity of the electrons, and, therefore, a frequent placement of these lenses would decrease of the effective cooler length.

Note that the contradiction between maintaining small angles and strong focusing in the cooling section is very different in the scheme with continuous magnetic field, because an electron propagates along the field line corresponding to its origin. If the field magnitude is changed in the entire beam line, the value of the angles may stay the same.

The decision was made to combine the advantages of these two schemes, lumped focusing and continuous magnetic field, by introducing a longitudinal magnetic field in the cooling section that is large enough to counteract possible perturbations but still low enough to allow beam transport outside of the cooling section with lumped elements and the use of a standard electrostatic accelerator with minimal modifications. An important requirement of this new scheme is the cathode immersion into a longitudinal field, so that the magnetic fluxes at the cathode and in the cooling section are matched [17].

Note that applicability of such scheme is critically dependent on the magnetic flux required in the cooling section. When a beam with no transverse velocities inside a solenoid exits into a free space, conservation of the canonical angular momentum results in a coherent angular rotation of the beam. In the paraxial ray approximation, it is equivalent to the appearance of an effective normalized emittance [17]

$$\varepsilon_{B,eff} = \frac{e\Phi}{2\pi m_e c^2}, \qquad (12)$$

where $\Phi = B_{CS} R_{CS}^2$ is the magnetic flux through the beam cross section in the solenoid. As in the case with a real emittance, the beam transport with lumped focusing is possible only if this emittance is sufficiently low. For example, let's consider a transport channel for $\gamma = 10$ with a typical beam radius of ~1 cm and a beta-function of ~1 m. If the required beam radius in a cooling section solenoid is $R_{CS}$ ~ 1 cm, Eq. (12) limits the solenoid magnetic field to ~300G. To use lumped focusing during acceleration (i.e. at lower $\gamma$), the magnetic flux should be decreased even further in comparison with this example by reducing both the beam size and the magnetic field strength in the cooling section (in the REC case, to $R_{CS}$ = 2 - 4 mm and $B_{CS}$ =100 – 200 G).

The specific choice of the beam line optics is presented in section 4.1.

## 3.2. Electron cooler setup description

A schematic of the electron cooler is shown in Figure 3, and its elevation views are presented in Figure 4.

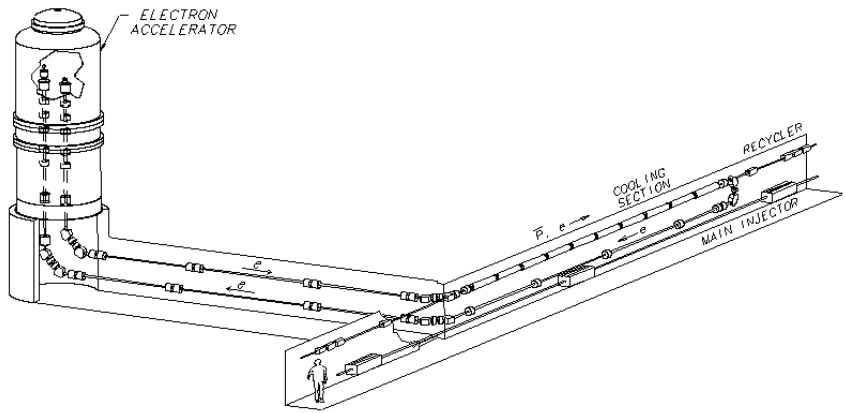

Figure 3. Schematic of the electron cooler.

Electrons are emitted from a thermionic cathode, accelerated inside an electrostatic accelerator, Pelletron [18], and transported through a beam "supply" line to the cooling section where they interact with the antiprotons circulating in the Recycler ring. After separation of the beams by a 180 degree bending magnet, electrons move through the "return" beam line, and back to the Pelletron through a "transfer" line. Inside the Pelletron, the electron beam is decelerated in the second ("deceleration") tube and is absorbed in a collector at the kinetic energy of 3.2 keV. The travel of the electrons from the emitter to the collector is called 'recirculation' in this paper, following a Fermilab colloquial convention. What it really entails is explained in detail in Section 5. Note that the Recycler shares the tunnel with a 150 GeV synchrotron, Main Injector (MI).

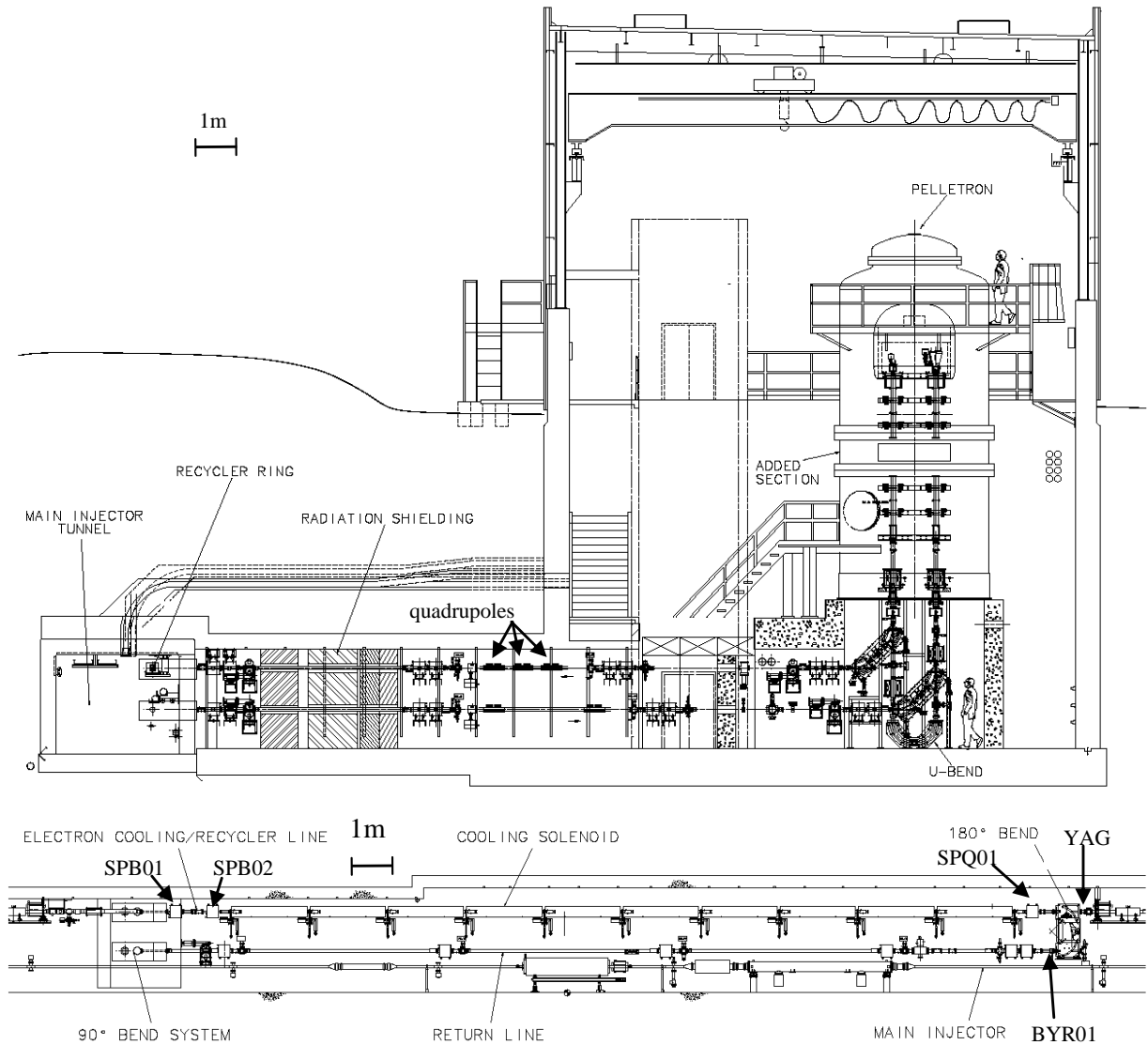

Figure 4. Upper - Elevation view showing the Pelletron and beam lines ("supply" and "transfer'). Lower - Elevation view of the portion of the Main Injector tunnel containing the cooling section and the "return" line.

When both main bending magnets under the Pelletron are turned off, the beam can be recirculated through a short beam line, denoted as U-bend in Figure 4. This so-called U- bend mode was used for commissioning purposes. For instance, in this mode we were able to reach DC beam currents of up to 1.8 A at the nominal energy.

The column of the nominally 6-MV Pelletron consists of 6 sections divided by hollow 1.9m diameter aluminum discs referred to as separation boxes (Figure 5). The space inside the separation boxes houses power generators driven by a mechanical shaft, focusing lenses, and electronics. The middle separation box is longer and also contains ion pumps for the acceleration tubes. The total length of each acceleration tube is ~3.6 m (12'). The high voltage is distributed along the column by a resistive divider. In addition, there are resistive dividers on each of two acceleration tubes. These three dividers are electrically connected at each separation box. All electronics is controlled through fiber optics.

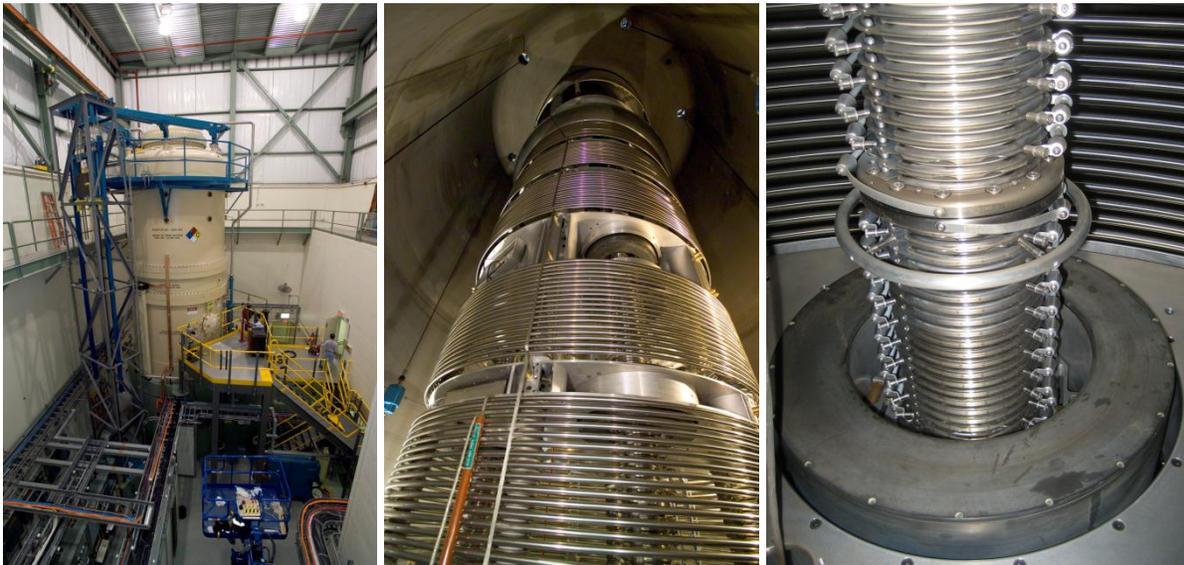

Figure 5. Photo of the Pelletron. Left – outside view; electronics racks are at the lower left corner. Center- view inside the Pellteron tank. The covers of the separation boxes (identifiable by the discontinuties in the aluminum field-shaping hoops) are removed, and the terminal shell is lifted (barely visible at the very top of the picture). During normal operation, the cables moving the elevator platform inside the tank are removed. Right – a view from inside the column looking at the deceleration tube and a lens (black annulus around the tube). On the lower left side is the fiber optics.

At the top of the column is the Pelletron terminal, where the electron gun and collector are mounted. The terminal has the potential of the gun and collector anode; some of control electronics is grouped at a platform that is electrically isolated inside the terminal and stays at the cathode potential.

While description of the final design of the gun and its performance has not been published, the main ideas for a low-halo gun are described in Ref. [19]. Mechanical schematics of the gun are shown in Figure 6, and Figure 7 presents a photo of the gun installed in the Pelletron terminal. The cathode is positioned in the central plane of a short solenoid that determines the value of the longitudinal magnetic field at the cathode. Identical solenoids on the gun and collector side protruding down from the terminal provide beam focusing at low energies.

Some specifics of the collector are mentioned in section 5.2. Its mechanical schematic and a photo are shown in Figure 8.

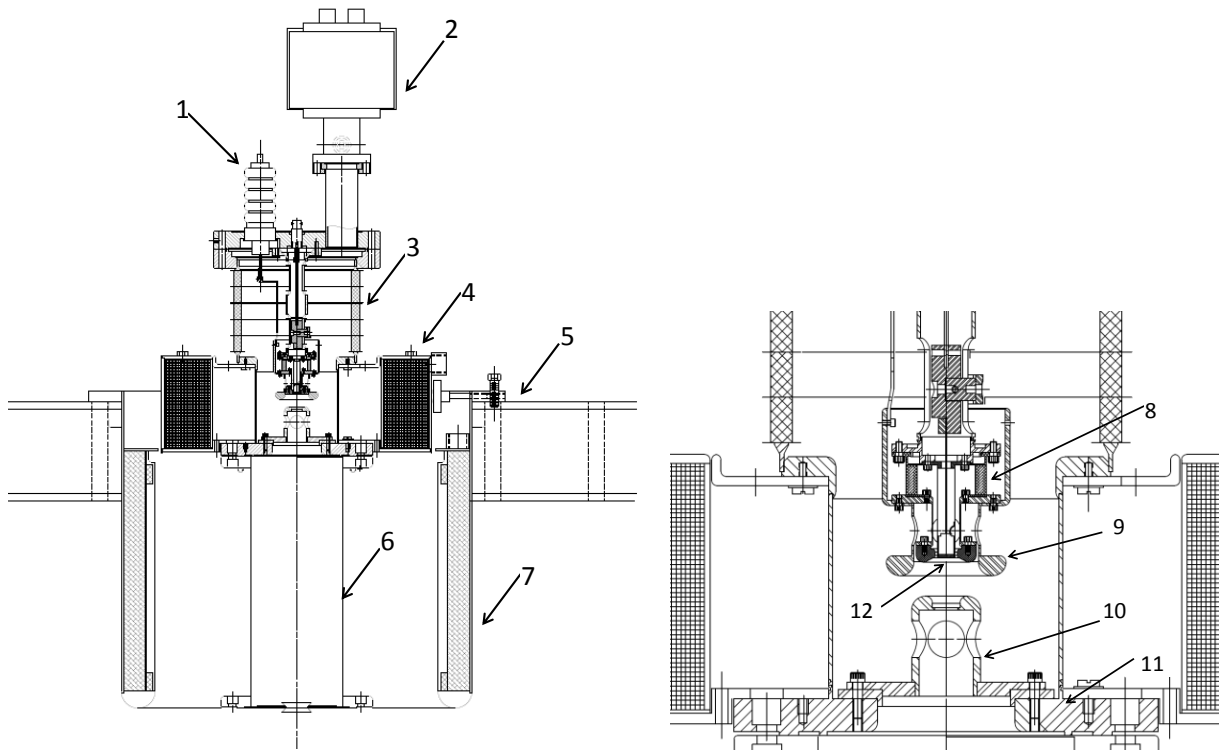

Figure 6. Mechanical schematic of the electron gun installed in the Pelletron terminal and a closer view of the cathode region. 1- control electrode feedthrough, 2- ion pump, 3- HV insulator, 4- short gun solenoid, 5- terminal platform, 6- the first acceleration tube, 7- long gun solenoid, 8- control electrode insulator, 9 – outside part of the control electrode (SS), 10- anode (SS), 11- gun flange (KOVAR), 12 – cathode and inside part of the control electrode (Hf).

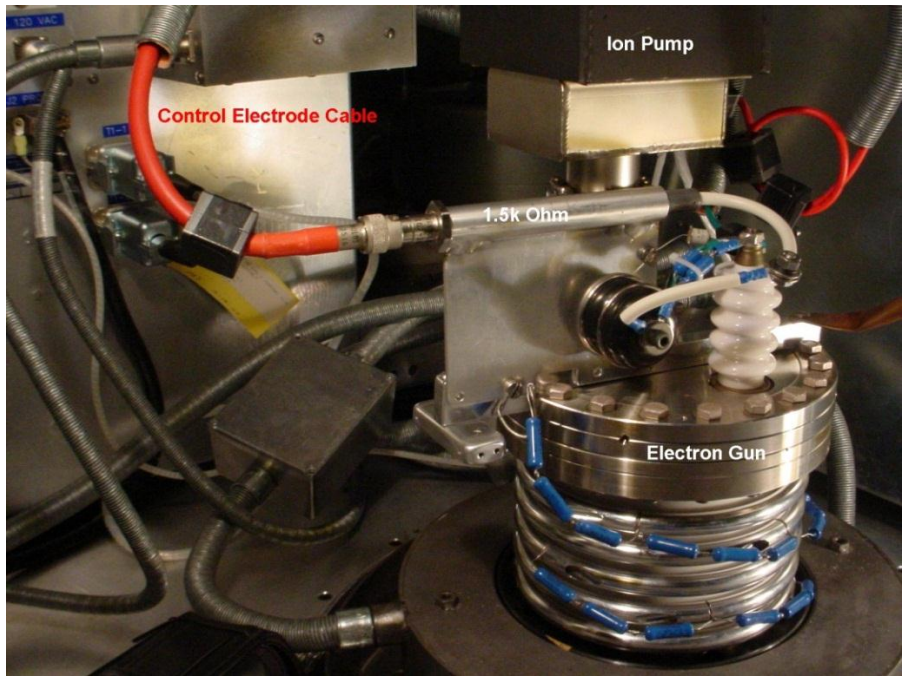

Figure 7. Photo of the gun installed in the Pelletron terminal.

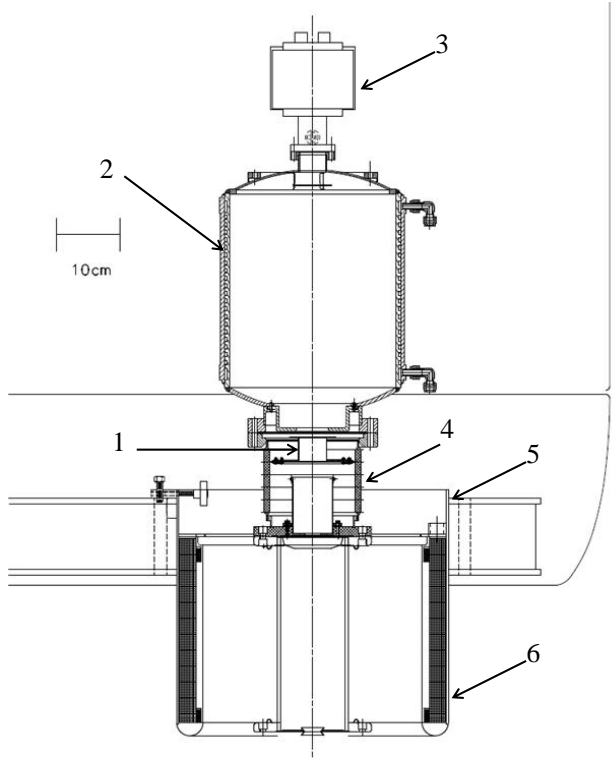
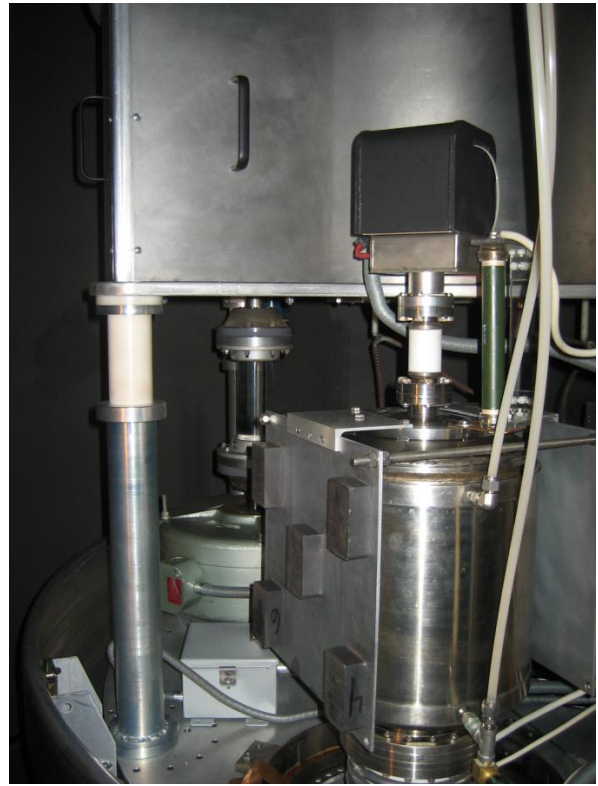

Figure 8. Mechanical schematic and photo of the beam collector installed in the Pelletron terminal. In the drawing: 1- Suppressor electrode, 2- water-cooled collector surface, 3- ion pump, 4- HV insulator, 5- terminal platform, 6 – collector solenoid. The steel plates and magnets creating the transverse magnetic field in the collector are not shown. Visible in the photo: the collector with steel plates and magnets (right); the generator (behind the collector) installed on the axis of the rotating shaft; a platform at the cathode potential (covered, top); an insulated part of the shaft (translucent cylinder above the generator) connected to another generator supplying power to electronics at the platform. The terminal shell is lifted.

Most of focusing is provided by solenoidal lenses. Inside the Pelletron, the lenses are mounted in each separation box; a pair of dipole correctors is incorporated into the body of each lens.

In the supply and transfer lines, the solenoids are grouped into doublets (an example is shown in Figure 9). Each doublet is fed by a single power supply, and the solenoids magnetic fields have opposite direction to eliminate the total beam rotation angle. It is a part of the optics design creating the rotationally invariant beam transport line between the Pelletron and the cooling section [20]. In the return line, focusing is made with single lenses of the same air-cooled type.

Correction of focusing is made by 6 weak, printed-circuit quadrupoles in the supply line that were used to optimize electron angles in the cooling section (see Section 6.2). Similar quadrupoles were installed in the transfer line and were intended to correct the shape of the beam entering the collector but were not used in operation.

All bending is done either in vertical or in horizontal planes. In most of cases, the beam is turned by 90º. The 90º bending assembly consists of two 45º zero-gradient sector magnets with a water-cooled solenoidal doublet in between to eliminate the dispersion after each bend in the supply line. Because of space limitations, no diagnostics was included into the 90º bend design, while the phase advance along the assembly was large (~2π). The consequence was that it was difficult to tune the bends and reconstruct the optics. The bending field ~300G was regulated using measurements from dedicated NMR probes (the system was designed and manufactured by Budker INP, Novosibirsk).

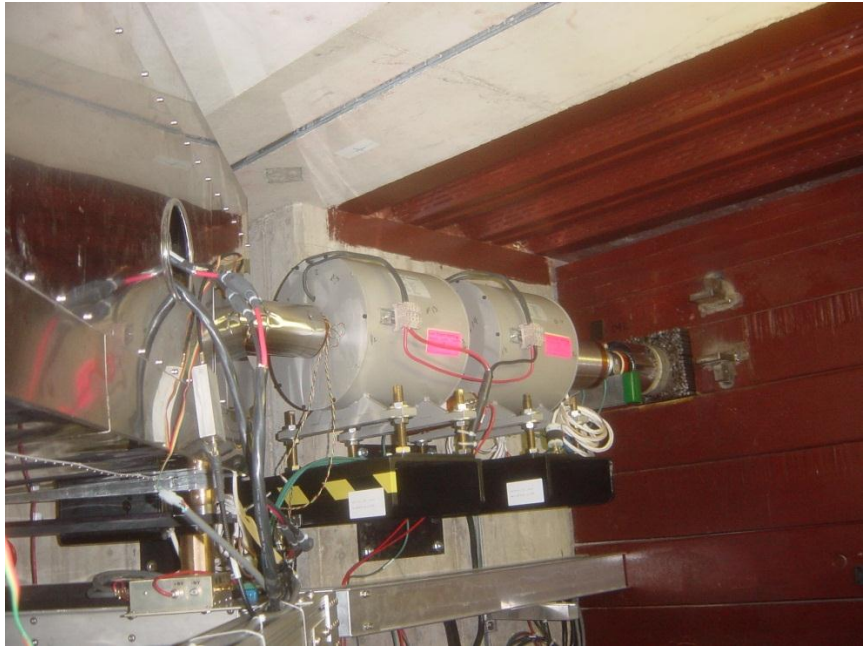

Figure 9. A doublet of solenoidal lenses in the supply line inside the MI tunnel. The red wall on the right is radiation shielding of a line toward the Pelletron. The shiny box on the left is the magnetic shield of the bend that merges the electron and antiproton beams. Parts of the beam line outside of the focusing elements are covered by µ - metal magnetic shield. The transfer line is below.

The short vertical separation between the cooling section and the return line forced a different design for the 180º bend that separates the electron and antiproton beams downstream of the cooling section and brings the beam into the return line (Figure 10). The bend consists of two 90º sector dipoles with index ½ and a quadrupole in between. The design allowed eliminating the dispersion while preserving the beam rotational symmetry. However, in operation it was found more beneficial to run with a large dispersion in the return line (see Section 5.3), and the quadrupole was kept off.

The vacuum chamber is pumped down by ion and titanium sublimation pumps. The typical diameter of the beam line vacuum chamber is 75 mm, while the beam aperture is limited by the BPM's inner diameter of 47 mm. The typical pressure was ~0.3 nTorr. The residual gas content was recorded mainly under the Pelletron, where the major components were ~80% hydrogen, ~10% CO/N2, and ~5% of water.

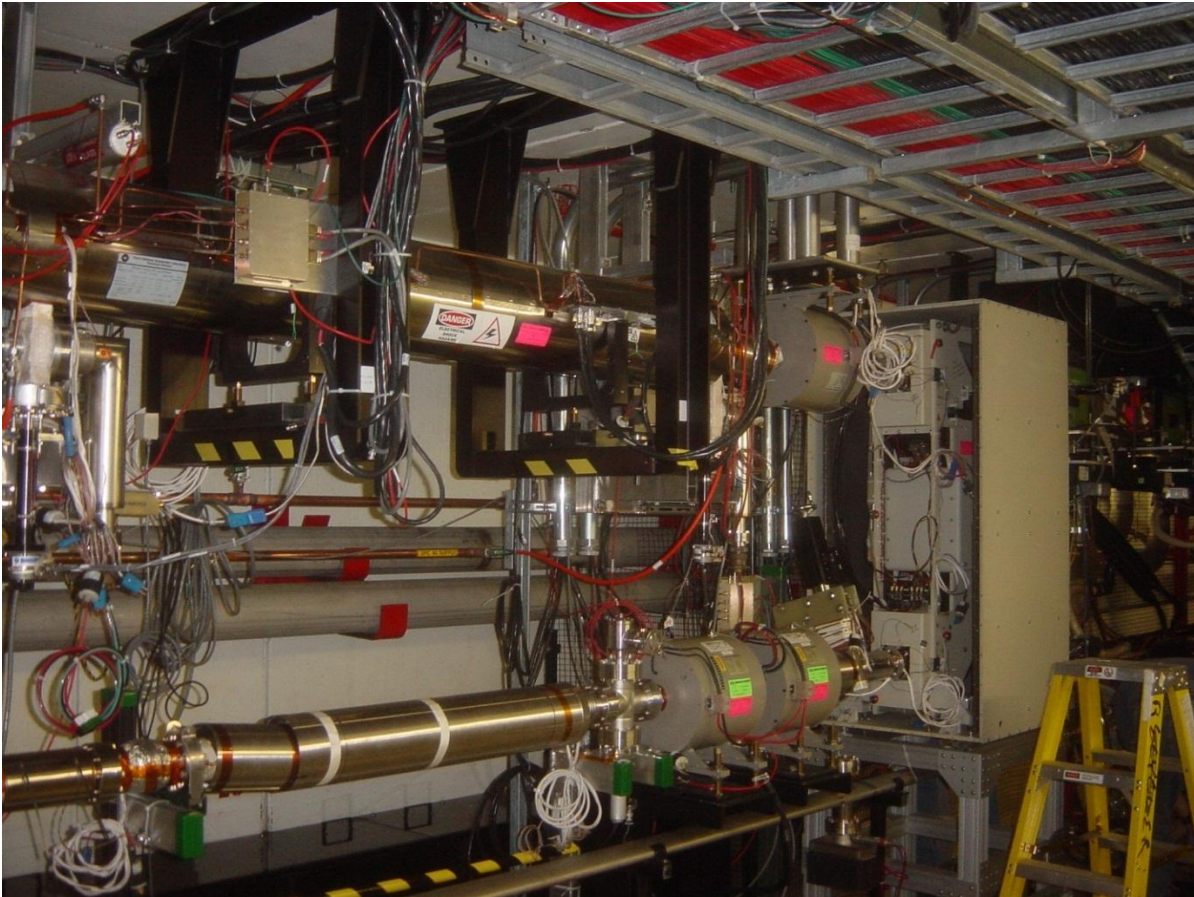

Figure 10. Photo of the 180º bend (far right) and the adjacent part of the cooler. At the time of taking the photo, the front part of the bend's magnetic shielding box was removed. Also shown are the last module of the cooling section (top) and the first solenoidal doublet of the return line (bottom). Most of the beam line is covered by magnetic shielding. The elliptical pipe under the return line is the Main Injector (MI) vacuum chamber; the black box along the wall shields magnetically the MI current busses.

### 3.3. List of the main points of concern for the design of the cooler

The following topics were identified as the most important to be resolved for successful operation of the electron cooler:
- Electron beam transport
- Stability of the beam recirculation
- Electron angles in the cooling section
- Energy ripple, stability, and absolute calibration

They are discussed in detail in the following sections.

## 4. Electron beam transport

### 4.1. Design

The beam envelope generated by the OptiM code [21] is shown in Figure 11a. The main features of the electron cooling line design are as follows:

1. Magnetic flux through the beam cross section at the cathode is equal to the flux at the cooling section (for a negligible space charge).
2. Nominally, the beam line between the Pelletron and the cooling section should be the rotationally invariant. Because the line includes not-rotationally-invariant elements, zero-gradient dipoles, it can be accomplished only for a specific tuning of the line.
3. Zero dispersion in the cooling section.
4. Possible rotation invariance and zero dispersion in the return line.

The first three conditions are needed to obtain and maintain low electron angles in the cooling section and the last one was considered to be useful for the beam transport inside the deceleration tube (but eventually was not used). The idea of the transport channel and its practical implementation in the cooler's prototype are described in detail in Ref. [20]. The final beam line differs from the prototype only by the length and the number of focusing elements.

Keeping the beam line after the cooling section nearly dispersion-free was successfully tested but found to be disadvantageous because it led to full discharges of the Pelletron (section 5.3). For an optimum protection from full discharges, the dispersion in the return line was made significantly larger than in the deceleration tube, ensuring that a beam with an energy lower than the nominal is lost well before reaching the deceleration tube electrodes.

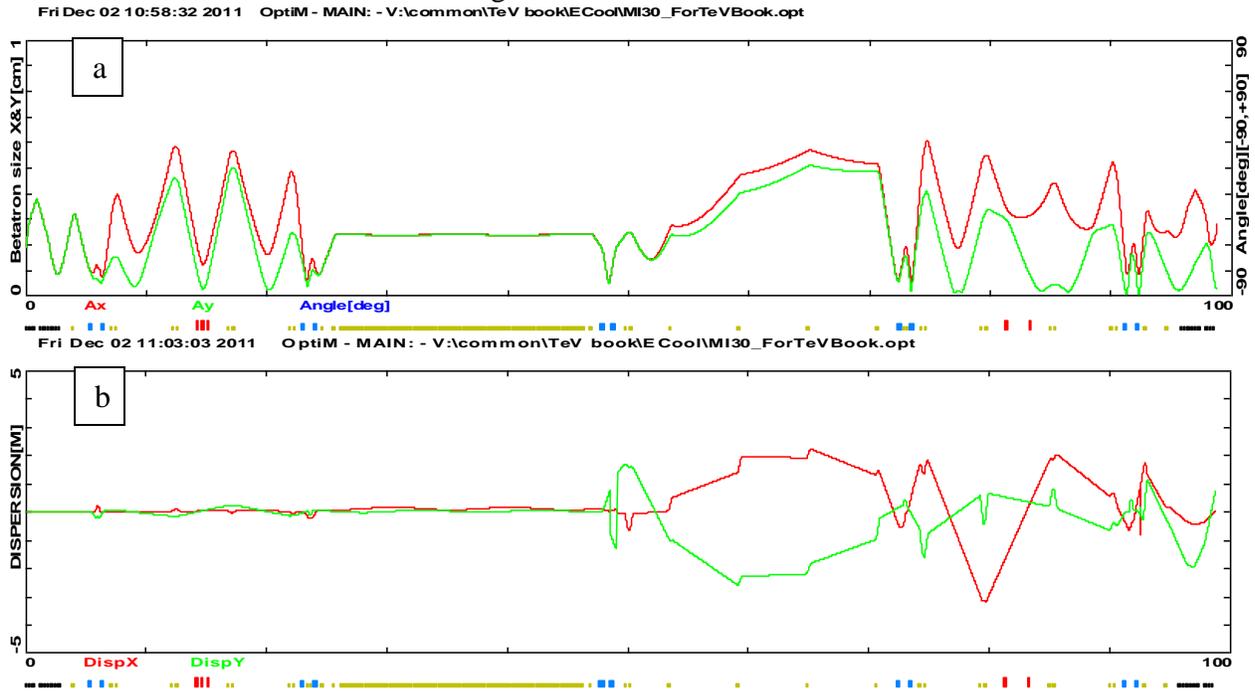

Figure 11. The electron beam envelope (a) and dispersion (b) simulated with the OptiM code. Settings of optical elements are as they were in regular operation, except for minor adjustments to the correction quadrupoles in the supply line. Initial conditions are adjusted to have a parallel beam in the cooling section. $I_e$= 0.1A.

## *4.2.  Commissioning of the beam line*

The relatively low momentum of the electrons makes the beam sensitive to various residual or fringe magnetic fields, which initially complicated the beam transport. In addition, calibration of the lenses' strength and knowledge of the initial conditions in simulations was found to lack the accuracy needed to pass the beam through without significant tuning.

Commissioning of the beam line was made first with a pulse mode. The gun generates 2 μs, 1 Hz pulses that are analyzed by the beam line BPMs demodulating at 130 kHz [22]. When the beam hits the vacuum chamber somewhere close to a BPM, the secondary electron shower creates a large signal at the BPM sum (intensity) output. The intensity was recorded as a function of the current of an immediately upstream corrector, and the current was set to center the distribution. An indication of the loss-free beam passage through a BPM was to observe a constant intensity over some interval of corrector currents and, correspondingly, of the beam transverse position within the BPM. Typically, this intensity was by several times lower than its peak when the beam would touch the BPM electrodes.

The most complicated tuning was inside the deceleration tube, where there are no BPMs. The main indicators were the beam loss at the bottom tube electrode, the peaks in the anode power supply current, and the total drop of the terminal voltage during the pulse.

After passing through the entire beam line a pulsed beam of ~0.5 A with losses not resolvable in the pulsed mode (<10 mA), a DC beam of several mA went through as well. Further tuning toward higher DC beam currents was made by measuring the beam loss as a function of the transverse position in various locations along the beam line, hence determining aperture restrictions.

The initial intent was to use the optics model for fine tuning of the beam angles in the cooling section. The optics was measured by deflecting the beam with dipole correctors, one at a time, or by changing the beam energy while recording BPM readings [23]. The resulting differential trajectories (i.e. the initial, unperturbed trajectory subtracted from the deflected trajectory) were analyzed with OptiM [21]. Multiple sets of these measurements allowed correcting electrical connections errors, polarities etc. as well as adjusting calibrations of the optical elements.

While the procedure permitted to correct the calibrations down to the level of several percent, more accurate results were not achieved. The variations between sets of differential trajectories measurements made with different correctors or days apart were well above the statistical BPM errors of ~ 10 μm. As a result, the agreement of the model with the measured trajectories was good only within relatively short portions of the beam line, but attempts to trace the entire cooler were not satisfactory. An example of such measurements and its comparison with the corresponding OptiM simulation is shown in Figure 12. The differences between the data and the simulations seen on Figure 12 were likely the result of mechanical drifts of the optical elements, insufficient stability of the power supplies, and non-linearity of the focusing fields.

Another complication for modeling the beam line is the accuracy to which the initial conditions are known. They were derived from simulating the electron gun with the UltraSAM code [24] and then propagating the simulation through the acceleration tube with the BEAM code [25] to the tube exit, where the OptiM optics file begins. The simulations were tested against measurements of the beam profile with an Optical Transition Radiation (OTR) monitor, which is mounted right below the first bend. While a good agreement was reported [26], resolution of the measurements might not be at the level needed to rely upon for fine tuning.

Finally, simulations did not take into account the effect of secondary ion accumulation, which makes a significant contribution to focusing (section 6.4).

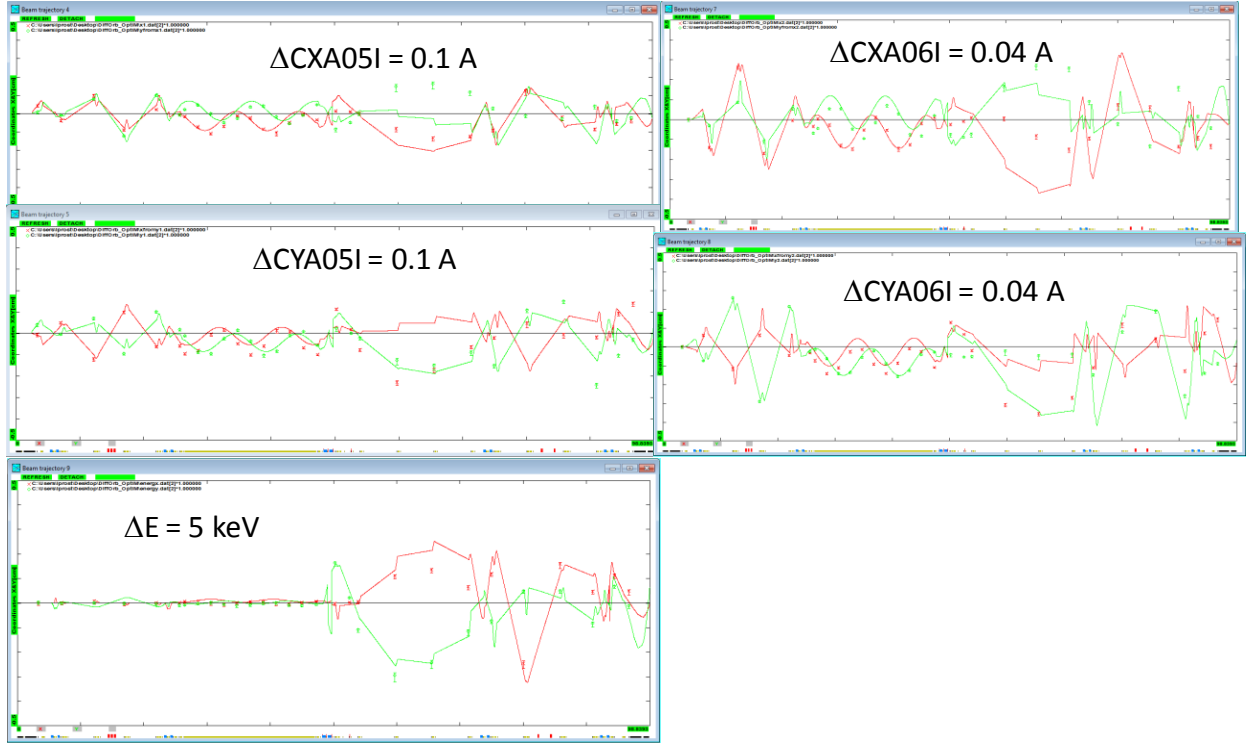

Figure 12. Example of a standard optics measurement. Points are the measured differential trajectories, and the solid lines are fits by OptiM with fudge factors adjusted from previous measurements. Upper four plots show responses to kicks by two pairs (X and Y) of correctors (last pairs in the Pelletron and first pair outside, before the first 90-degree bend), and the lower plot is the result of an energy increase. All trajectories are differential, i.e. shown after subtraction of the unperturbed trajectory. Red and green curves represent X and Y projections, correspondingly. The error bars for the data points are the statistical errors of the measurements.

Not knowing the optics with good precision, tuning of focusing within the cooling section was made by using of a set of 11 scrapers with round openings [27]. With the electron beam off, one scraper is inserted into the beam path, such that when the beam is reestablished, it goes through the scraper hole. Then, the beam was moved with dipole correctors in 8 directions until ~$10^{-5}$ portion of the beam was scraped each time, which was used as a definition of the beam boundary touching the scraper. With known calibration of correctors and openings diameter (15 mm), the measurements gave the dimension of the beam in 4 directions, and an ellipse fitted to these 4 numbers was assumed to be the beam shape. Repeating the procedure in all scrapers one by one gave the beating of the beam ellipse along the cooling section. While by design this diagnostics allows only to align the beam halo, which angles can differ significantly from the core [28], it was sufficiently good to demonstrate cooling and put the cooler into operation. All further tuning was made using cooling itself as an indicator of the electron beam angles.

## 5. Stability of the beam recirculation

### 5.1. *The energy recovery scheme and beam loss limitations*

To keep the dissipated energy low while using a MW-range DC beam in the cooling section, the cooler employs the energy recovery scheme. After acceleration and interaction with antiprotons, electrons return the energy to the terminal by decelerating in the second Pelletron tube down to the energy of ~3 keV at the collector surface and flow through the collector power supply toward the cathode to repeat the journey. At Fermilab this process is called 'beam recirculation'. A simplified electrical schematic is shown in Figure 13.

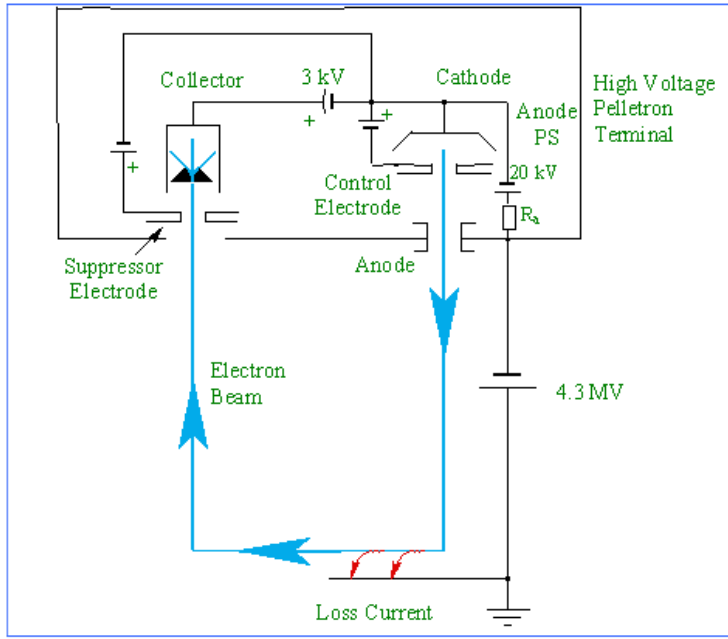

Figure 13. Simplified schematic of beam recirculation.

This scheme puts stringent limitations on the beam loss. The most obvious reason is the low current provided by the Pelletron chains (nominally up to 400 µA), which is by several orders of magnitude lower than the beam current. However, the beam loss inside the Pelletron tubes is restricted even more. A loss comparable with the current flowing through the tube resistive divider (~40 µA) significantly redistributes the potential along the tube. The resulting change in the beam envelope usually causes even larger losses, and the beam recirculation is lost in a matter of milliseconds.

However, for stable long-term operation much lower losses in the tubes are required, at the level of several µA. We interpret this as a result of a charge accumulation on the tube ceramic and following partial discharges in the acceleration gaps. These discharges occur all the time, with frequency dependent on the tube voltage gradient and amount of the beam loss. The structure of the Pelletron column contains large aluminum discs called separation boxes, which are connected every ~60 cm (2') to both tubes resistive divider as well as to a dedicated column resistive divider. When only one of 42 gaps contained between neighboring separation boxes is discharged, the effect on the voltage outside this portion of the tube is negligible. Hence, by itself, a discharge of a single gap cannot significantly change the overall voltage distribution. However, with some probability the plasma formed from such discharge can shorten several neighboring gaps, while the capacitance between the separation boxes still holds the total voltage constant. If the unaffected portion of the tube is capable of holding the entire voltage, the gaps charge up again, and the recirculation is not interrupted. Also, if the beam envelope modification that results from the temporary charge redistribution is large but lead to beam loss only somewhere outside of the Pelletron, the protection system interrupts the beam and normal operation can be restored in a matter of seconds (so-called "a beam trip"). Otherwise, the entire tube shortens, and the Pelletron voltage drops to nearly zero. Dealing with this "full discharge" scenario is described in section 5.3.

## 5.2. Steps to limit beam losses to the acceleration tubes

Because decreasing the beam loss to the tubes was recognized as the only way to provide long uninterrupted recirculation, several distinctive steps were made at the design and R&D phases.

- Electron gun with a negatively biased 'control electrode'.

Experience of MeV-energy electrostatic electron accelerators showed that one of the problems is to turn on the beam, because in the absence of a strong longitudinal magnetic field the beam may strongly diverge at low currents. It was alleviated by employing a gun with an electrode near the emitter that is negatively biased with respect to the cathode (aka the control electrode) [19] and shapes the electric field near the emitting surface. Thus, in this gun, the emitting area is determined by the location of the zero equipotential surface at the cathode that results from the relative difference between the anode and control electrode voltages. Hence, for a small beam current, only electrons from a narrow portion of the emitter near the axis are accelerated. As a result, the beam diameter and divergence increase monotonically as a function of the beam current until the entire cathode surface emits, at which point the beam size and envelope angle roughly stabilize. Consequently, a focusing channel optimized for the nominal current can transport lower currents as well.

An equally important feature of this gun in its use of a control electrode is the suppression of electron emission from the side surface of the emitter. With a positively biased control electrode, such electrons are accelerated and create a large beam halo, increasing the beam loss.

- Effective electron beam collector.

Another source of the current loss is electrons from the beam collector region. First, part of secondary electrons created at the beam absorbing collector surface escapes and is accelerated in the deceleration tube. Development of a collector with a transverse magnetic field [29], [30] allowed decreasing this portion to less than $10^{-5}$ of primary beam current. The second phenomenon is intra-beam scattering that increases the energy spread in the beam, causing its low-energy tails to be reflected while decelerating near the collector [31]. This effect determines the minimal value of the collector potential and limits the applicability of the traditional method of decreasing the secondary electron flow from a collector by creating a potential minimum near the collector entrance. It was taken into account when designing the final version of the collector, where the collector efficiency for a nearly monoenergetic beam on a test bench weakly depended on the potential of the electrode near the collector entrance. In operation, this so-called "suppressor" electrode was at +3 kV with respect to the cathode while the collector voltage was 3.3 kV.

- Optimum beam tuning

We realized that tuning is optimum for the stability of operation when the beam loss to the acceleration/deceleration tubes is minimized, which not necessarily coincides with the minimum of the overall beam loss. In particular for the deceleration tube, it meant passing toward the ground as many as possible of the electrons escaping from the collector, even at the cost of increasing the total beam loss. Typically, the changes of the tube resistive dividers currents after turning the electron beam on were used as an indication of a beam loss to the tubes.

- Suppression of the electrons emitted from the control electrode
  While the gun design suppresses the beam halo from electrons emitted from the cathode, particles emitted from the control electrode have a similar energy as the main beam and may propagate out of the gun. However, because the initial conditions for these particles are very different from those of the main beam, such particles are lost in the acceleration tube, deteriorating the long-term stability of the beam recirculation. This effect is difficult to model on a low-energy test bench, and all the significant stability improvements were made as results of experiments with the Pelletron.
  First, the recirculation stability dramatically improved when the aperture in the gun anode was decreased to the minimum size allowed by the envelope of the primary beam [32]. With the smaller aperture, most of the electrons emitted from the control electrode are lost at the anode.
  An important mechanism for generating electrons from the control electrode is through the impact of secondary ions. Without special precautions, the ions created by electrons in the beam line are captured by the beam space charge and can travel along the beam toward the acceleration tube, where they are accelerated and end up in the gun, irradiating, in part, the control electrode. This mechanism was suppressed by creating a potential barrier at the entrance of both tubes with a positive voltage on the corresponding BPM plates, which increased the typical time between beam interruptions from dozens of minutes to hours.
  Operationally, the recirculation stability was found to deteriorate when the pressure in the tubes was rising. To keep a typical interval between beam trips above several hours, the first ion gauge below the tubes had to be kept below ~0.3 nTorr. We attribute this to the same effect as in the previous paragraph, i.e. to electrons generated by impacts of secondary ions, but in this case, by ions generated inside the tubes.
  Finally, unwanted electrons were found to be emitted also from the inner surface of the negatively biased control electrode [32]. Following the hypothesis that the emission is related to the cathode material being sputtered onto the control electrode by ions striking the emitter (primarily during full discharges), we looked for a material of the control electrode that would suppress the emission from barium films. Replacing the originally copper electrode by tantalum and then by hafnium (the part marked 12 in Figure 6) solved the problem.

## 5.3. Full discharges

The most destructive events in operation are full discharges, when the Pelletron voltage drops to nearly zero in a matter of a microsecond. This occurs when a significant portion of one of the acceleration tubes is shorted out by the formation of plasma in the vacuum before the resulting increase of the voltage gradient on other parts of the tube causes the protective spark gaps on the gas side to fire. Currents flowing in vacuum modify the electrodes surfaces, decrease the electric strength, and create a large burst of pressure (to 0.01 mTorr), which may take hours to recover from. Also, much higher currents flowing on the gas side produce large electromagnetic waves, damaging the equipment inside and sometimes even outside of the Pelletron. Full discharges were common during the R&D and commissioning phases, and significant efforts were put to decrease their frequency.

Two factors were recognized to be of primary importance to avoid full discharges: a high electric strength of the tubes and preventing the primary beam from reaching the tube electrodes.
- At the R&D phase it became obvious that the electric strength of the tubes has to exceed significantly the nominal potential gradient. At that time the total length of the accelerating/decelerating tubes was ~3 m (10'), divided into 5 two-foot long sections. Each section was conditioned individually to ~ 1.1MV (without any apparent discharges for many minutes after conditioning) reaching occasionally a maximum of ~1.3 MV. With this conditioning procedure the full length tubes were able to hold ~5 MV without beam (nominal is 4.3 MV), but at the same time operation with beam was stable only at 3.5 MV, ~60% of the sum of the maximum voltages of individual sections after they had been conditioned. A dedicated experiment showed that the factor of 0.6 remains roughly the same if a portion of the tubes is shorted. As a result, the decision was made to add one more two-foot long section in the final Pelletron assembly. In accordance with the pattern just described, raising the total tube length to 3.6 m eventually allowed stable operation at 4.3 MV.

- Even if the tubes electric strength is appropriate, a full discharge is very likely when the beam core touches the tubes electrodes [33] because a large portion of the tube can be affected simultaneously. Excluding studies and tuning, such situation is possible when there is a sudden increase of the beam loss anywhere along the beam line. Because the lost current discharges simultaneously the terminal and the cathode-anode effective capacitances, both the Pelletron and gun voltages drop down, changing the beam envelope and, at locations with high dispersion, the beam trajectory. Initially, such changes of the beam envelope in the acceleration tube were the leading cause of the full discharges. This was greatly suppressed by optimizing focusing such as to pass the beam through the acceleration tube for a range of gun parameters as wide as possible. On the other hand, the deceleration tube was protected by creating a large dispersion in the return line while minimizing it in the tube.

- Another critical element to fight the full discharges was the implementation of a fast protection system [34], closing the gun in about 1 μs after detecting a drop of the terminal voltage or a high beam loss (through radiation monitors). By the time of shutting the gun down, the typical terminal voltage drop was 5-10 kV, and corresponding beam envelope modifications were not large enough to irradiate the tube electrodes.

With all these elements in place by the fall of 2005, operation at the beam current of 0.5 A was producing one full discharge every several days on average. Switching to 0.1A for normal operation decreased the frequency of the full discharges dramatically, to a few per year (excluding studies and cases with broken equipment).

### 5.4. *Beam trips*

During regular operation, there are typically several beam trips per day. A trip starts with the protection system detecting either a drop of the terminal voltage or an elevated radiation near the beam line. Then an analog circuitry quickly shuts the beam off by applying a large negative voltage to the gun control electrode, and a software loop, inherently slower, turns off the anode voltage and disables the gun operation permit as an extra layer of protection. Following the event, another software loop checks the Pelletron voltage and vacuum, and if they are within their tolerances, restores the electron beam in about 30 seconds after the beam trip.

The majority of the trips can be associated with one of the following reasons: false trigger due to Main Injector beam losses or from excessive noise from the electronics reading the terminal voltage; high beam loss near the entrance of the deceleration tube due to an electron beam trajectory alteration (for example, caused by a large variation of the fringe fields of the Main Injector); a partial discharge in one of the tubes. While reasons for some of the trips are unclear, the negligible impact of the interruptions on operation at this frequency did not justify additional efforts to understand them.

# 6. Electron angles in the cooling section

The value of the electron angles in the cooling section dramatically affects the cooling properties. For typical operational parameters, the rms transverse electron velocity $\sigma_{ex}$ (in the beam frame) significantly exceeds the antiproton velocity and the longitudinal electron velocity component. For this case, Eq. (11) predicts that the longitudinal and transverse cooling rates (time derivatives of the emittances) depend on the angles (or transverse velocities) as $\sigma_{ex}^{-2}$ and $\sigma_{ex}^{-3}$, correspondingly. Therefore, cooling efficiency is very sensitive to variations of the angle value.

The origins of the angles can be roughly divided into four categories:
- incoherent angles originated from the thermal electron velocities at the cathode
- angles resulting from an envelope mismatch
- nonlinearities in the beam line
- coherent dipole motion

The following sections describe these effects in detail. To give a sense of scale, let us note in advance that the estimated 1D effective electron angle was in the best cases ~100 μrad.

## 6.1. Thermal angles

As it was noted in section 4.1, envelope matching requires equality of the magnetic fluxes through the beam at the cathode and in the cooling section (neglecting the space charge)

$$B_{cath} R_{cath}^2 = B_{CS} R_{CS}^2 , \quad (13)$$

where $R_{cath}$ and $R_{CS}$ are the beam radii and $B_{cath}$ and $B_{CS}$ are the magnetic field strengths at the cathode and the cooling section, respectively. Ideal matching means that no additional coherent motion is excited, and the incoherent transverse momentum at the cathode with temperature $T_{cath}$, $p_{Tcath} = \sqrt{m_e k_B T_{cath}}$ leads to an increase of the transverse momentum in the cooling section

$$p_{T\_CS} = p_{Tcath} \sqrt{\frac{B_{CS}}{B_{cath}}} . \quad (14)$$

The corresponding 1D electron angle is

$$\alpha_T = \frac{1}{\gamma\beta} \sqrt{\frac{k_B T_{cath}}{m_e c^2} \frac{B_{CS}}{B_{cath}}} . \quad (15)$$

For the cooler parameters, $T_{cath} = 1050\,°C$, $B_{cath} = 86\,G$, $B_{CS} = 105\,G$, Eq. (15) gives a 1D thermal contribution of $\alpha_T = 57\,\mu rad$.

## 6.2. Envelope mismatch

An envelope mismatch (a focusing error) results in electron angles linearly growing with radius and the beam shape and size changing along the cooling section. Initial tuning was made by measuring the envelope with scrapers [27] (see section 4.2). The beam envelope was found to be close to being round and was adjusted with two lenses right upstream of the cooling section. The residual 2D angle at the beam periphery was reported to be 0.22 mrad [28]. This was sufficiently good to obtain electron cooling rates adequate to demonstrate cooling and early operation. Attempts to further adjust the lenses according to the cooling rates did not lead to any noticeable improvements.

However, there were accumulating indications of a deficiency to the procedure. The beam boundary was determined by scraping ~$10^{-5}$ of the beam intensity. Therefore, this procedure was sensitive only to the beam halo, which properties may be very different from the core's [28]. At the very beginning, the measured beam size exceeded the prediction from Eq. (13) by a factor of 1.3, well above the measurement errors. Later, comparison of the cooling force measured on the beam axis and the cooling rates predicted by Eq. (11) showed a significant discrepancy that could be resolved only with the assumption of a large envelope mismatch [35]. Finally, direct imaging in a pulse mode with a scintillator located at the end of the cooling section clearly revealed that the beam core was elliptical [36], which would explain the relatively poor cooling rates. By adjusting quadrupoles upstream of the cooling section, the beam core was properly matched in the pulse mode but it did not improve cooling, likely because the residual ion background in the DC mode significantly modified focusing (see section 6.4). There was no non-destructive transverse beam size diagnostics available to tune the beam envelope, and the accuracy to which the beam optics was measured did not allow using simulations for fine tuning either.

Eventually, the quadrupoles were tuned based on the cooling properties. The drag rate was measured (see the discussion of the procedure in section 8.1) for several current settings of one of the quadrupoles, which would then be set to the value giving the highest cooling force. Going through this procedure for all quadrupoles successively noticeably improved the cooling properties at the beam current $I_e = 0.1A$. For example, at 1 mm offset the drag rate increased by a factor of ~2 [36]. Estimation made with OptiM showed that the applied changes to the quadrupole currents should correct the angles of a 0.1A electron beam at the radius of 1 mm by ~0.15 mrad (2D).

Note that this tuning procedure was not very reliable because a response in the drag rate to a typical change in the quadrupole current was comparable to the measurement scatter, and attempts to improve even further were not successful. Comparing simulations with different sets of quadrupole currents that were giving similar drag rates, one can speculate that the remaining envelope angle was about a half of what had been corrected. Assuming a linear dependence of this angle with the beam radius, the rms 1D contribution average over the beam cross section is estimated to be ~50 μrad.

An additional source of angles linear with the radial offset, which exists even for ideal matching, is the beam rotation caused by the electron space charge in conjunction with the longitudinal magnetic field $B_{CS}$ in the cooling section. For $I_e = 0.1A$, an electron beam radius $a_e = 2mm$, and $B_0 = 105G$, the contribution of this drift angle at the beam periphery is negligible with respect to all other sources:

$$\alpha_{dr} = \frac{E_{sc}}{\beta\gamma^2 B_0} = \frac{2I_e}{\beta^2\gamma^2 B_{CS} a_e c} \approx 10\,\mu rad \quad (16)$$

### 6.3. Non-linear perturbations

Several effects may result in a non-linear perturbation of the electron motion: space charge of the electron beam itself, higher-order field components in the focusing elements, and the electric field of the background ions.

The angles caused by the beam space charge are likely a minor contributor to the cooling properties. At the standard operating anode voltage of 20 kV and the beam current of 0.5A, for which the gun was optimized, the current distribution at the cathode is close to being flat (Figure

14) i.e. uniform density with sharp edges. Because the beam envelope is determined mainly by the effective emittance Eq.(12), the distribution remains nearly flat in the beam line as well, hence the force due to space charge is mainly linear. Measurements in the pulse mode demonstrated that the beam distribution was indeed flat [26]. However, for lower currents, the edges of the beam current density profile soften. Consequently, the non-linear component of the space charge, which arises from the rounded current density profile, increases while the linear component diminishes. Estimations show that for the operational current $I_e = 0.1$ A and focusing tuned to minimize the angles near the beam axis, the additional angle due to the current density profile at a radius of 1.8 mm is ~30 µrad, insignificant to the total angle in comparison with other contributions at this offset.

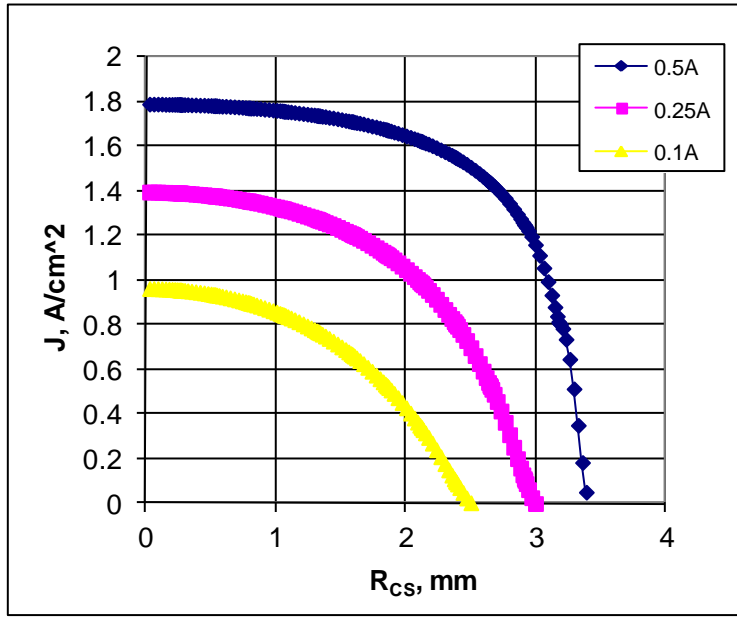

Figure 14. Current density distribution in the cooling section. The current density distribution was simulated at the cathode with the code UltraSAM [25] and then adjusted according to Eq. (13). The anode voltage was 20 kV. The values of the beam current are indicated on the plot.

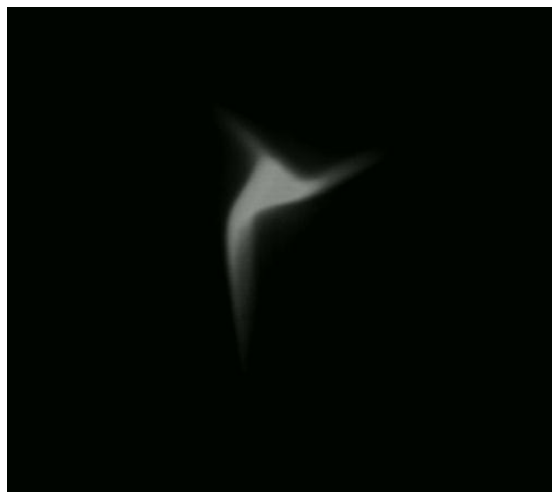

Figure 15. Image of the beam tightly focused at the YAG.

The next effect, a non-linearity originating from the imperfection of the focusing elements' magnetic field, became obvious during measurements of a pulsed electron beam imaged onto a YAG crystal installed right downstream of the cooling section. When the beam was tightly focused at the YAG, the image clearly showed higher-order perturbations (Figure 15).

To determine the source of the perturbations, the beam positions in the BPMs were measured as functions of the currents in various correctors. Then, the angles to the beam centroid generated from the lenses at different offsets were calculated and analyzed [37]. For an ideal, optically thin solenoidal lens, this angle $\alpha_{ideal}$ increases with the offset $r$ as

$$\alpha_{ideal} = \frac{r}{F_{ideal\_0}}\left(1 + k_{lens} r^2\right) \tag{17}$$

where $F_{ideal\_0}$ is the focusing length for a paraxial trajectory and $k_{lens}$ is a coefficient determined by the lens geometry. While the angle dependence on the offset reconstructed from the measurement differed from the function of Eq.(17), one can still characterize the deviation from linearity in a similar manner, calculating the coefficient $k_{lens}$ for a radius typical for the measurements (~1 cm). For most of lenses, this coefficient was found to be within a factor of two from the one found in simulations ($k_{ideal}$ ~3·10$^{-3}$ cm$^{-2}$). However, in several cases these deviations were significantly higher. Most notably, one of the lens doublets in the supply line (i.e. between the first vertical and the first horizontal bends), called SPS03, showed the coefficient being ~6 times higher than $k_{ideal}$. For optimum steering (i.e. trajectory where non-linearity is minimal) and typical beam sizes in this lens predicted by OptiM simulations for beam currents of 0.1A and 0.5A (i.e. ~3 mm and ~5mm radius, respectively), the additional electron angle in the cooling section at the beam periphery is estimated to be ~40 μrad and ~100 μrad, correspondingly. While no complete simulations were made, taking into account aberrations in all lenses together with beam misalignments would likely double these numbers. Note, however, that because of strong (approximately cubic) dependence of these perturbations on the offset, the beam core should not be dramatically affected. The estimation for the 1D rms angle for a well-aligned 0.1A electron beam gives ~20 μrad. In operation, the beam trajectory was periodically re-aligned to keep the beam center within 1 mm from the lens axes.

In fact, the strongest non-linear contribution to the total beam angle comes from the residual gas ions. This effect is described in the next section.

### 6.4. Effect from the ions generated by beam-background gas interactions

Ions created in the beam line by the electron impact on the residual gas molecules can significantly modify its focusing properties. The initial kinetic energy of the secondary ions is close to thermal, and the electric field of the electron beam prevents ions from escaping radially starting from mA-range currents. With no ion clearing mechanisms, the ion density would increase until reaching the electrons' (i.e. up to the neutralization factor $\eta$ ~100%). At $\eta$ ~100% the focusing effect from ions is by a factor $\gamma^2$ ~100 higher than defocusing from the beam space charge. Electron beam envelope simulations with the OptiM code indicate that the electron beam space charge becomes important in the beam line at the beam current $I_e$ ~ 0.1 A. Therefore, for the operational range of 0.1– 0.5 A, the effect of ions should be significant at $\eta$ ~1%, thus requiring effective ion clearing.

To address this requirement, all capacitive pickups monitoring the beam position in the cooler (BPMs) have a negative DC voltage offset on one of their plates, while the other plate is DC grounded. The resulting electric field prevents the formation of a potential minimum inside the pickup and removes ions in the vicinity of each BPM. To further estimate the process of ion accumulation, we assume the residual gas to be hydrogen at 0.3 nTorr. The calculated time for reaching η ~1% is ~200 ms. It is much longer than the time for a thermal – velocity $H_2^+$ ion to fly ~5 m between two neighbouring BPMs, ~3 ms, and, therefore, clearing with an electric field in BPMs should be effective. However, significant size variations of both the electron beam and the vacuum pipe along the beam line create local potential minima that prevent ions from travelling to the clearing field in the BPMs.

Also, solenoidal lenses providing focusing in the beam line are additional barriers for ions. Because the electric field inside the electron beam is primarily radial, the transverse component of the ion velocity is typically much higher than the longitudinal. Due to the conservation of the transverse adiabatic invariant, even the modest magnetic fields of the lenses (≤ 600 G) can confine the ions, further favouring a steady state concentration. The initial expectation was that the focusing effect of the ion background would be mainly a linear tune shift and, therefore, could be compensated by adjusting the lens settings.

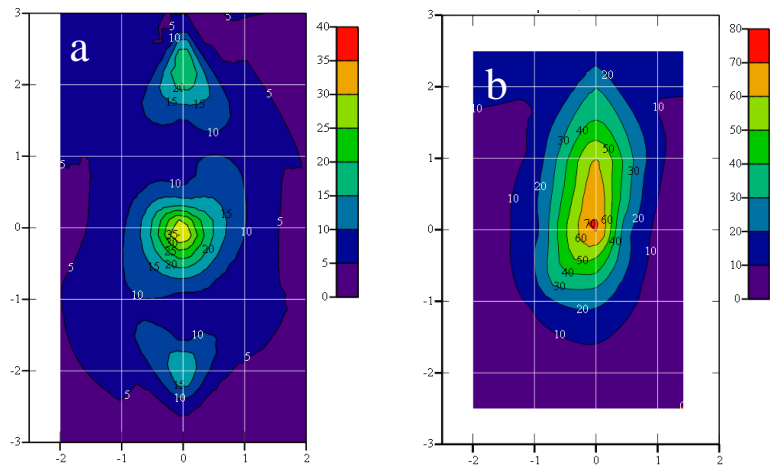

Figure 16. Contour plots of drag rates across the electron beam without (a) and with (b) ion clearing by beam interruptions. Voltage jump of 2 kV, $I_e$=0.3A. In the mode with ion clearing, the beam interruption frequency was 100 Hz. Contour levels are in MeV/c/hr. Horizontal and vertical axes are the corresponding beam displacements in mm. Note the difference by a factor of 2 in scales for the color bars. The data were taken on June 12, 2009 (a) and December 31, 2010 (b).

The cooling properties of the electron beam were found satisfactory for what became the standard operation mode, at $I_e$ = 0.1 A. However, the cooling efficiency (characterized by drag rate measurements) peaked at 0.1 – 0.2 A (curve 6/13/2006 in Figure 26), while it is supposed to be monotonically increasing with the electron beam current. Note that adjusting quadrupoles upstream of the cooling section (see section 6.2) significantly improved cooling at 0.1A but did not noticeably change its performance at higher currents.

A hint that it is related to the ion background came from results of transverse scans of the drag rates. A detail discussion of the drag rate measurements follows in section 8; here we assume that the drag rate is equal to the longitudinal cooling force averaged over a small-size ("pencil") antiproton beam. The drag rates were measured in various positions of this pencil beam with respect to the electron beam (experimentally, the electron beam was moved). Because the cooling force is determined mainly by the local properties of the electron beam and the value of the voltage jump is chosen to measure the drag rate near the maximum of the force curve of

**Error! Reference source not found.**, the force is roughly proportional to the local value of $j_e / \vartheta_e^2$. The results of such measurements in 31 points are shown as a contour plot in Figure 16a. With only three narrow areas providing significant drag rates, this profile corresponds to high-order focusing perturbations that cannot be corrected by adjusting solenoidal lenses and quadrupoles.

The only remedy to decrease the average ion concentration was found to be periodic interruptions of the electron beam. In the potential well created by the electron beam, ions gain the kinetic energy of up to 10 eV (at $I_e$ = 0.3 A). Thus, if the electron beam is abruptly turned off, an $H_2^+$ ion reaches the vacuum pipe in 1-2 μs. The capability of interrupting the electron current for 1 -30 μs with a frequency of up to 100 Hz was implemented in the electron gun modulator in 2009 [38]. While the clearing voltage applied to the BPMs was always on and certainly decreased the ion density, below for brevity we will refer to the operation with these interruptions as to the 'mode with ion clearing'.

Ion clearing significantly increased the area of the electron beam cross section with good cooling (Figure 16b) as well as improved the drag rate measured on axis at higher electron currents (Figure 26, curve 1/2/2011). The latter is related to the finite transverse size of the "pencil" antiproton beam in the measurements (see section 8.2).

Dependence of the drag rate measured at 1mm offset on the interruption frequency is shown in Figure 17. The results can be compared with the following model:

a. The beam space charge tune shift outside of the Pelletron tubes is relatively small, so that the envelope electron angle in the cooling section changes linearly with variation of the beam current and the offset in the beam $r$, $\Delta \alpha = k_{sc} \cdot r \cdot \Delta I_e$. According to OptiM simulations, the coefficient $k_{sc} \approx 1$ rad/A/m.

b. The envelope electron angle caused by accumulated ions is similar to the effect due to the beam own space charge,

$$\Delta \alpha_i = k_{sc} \cdot r \cdot \Delta I_e \cdot \delta \cdot \eta \cdot \gamma^2 \qquad (18)$$

where $\delta < 1$ is a fitting coefficient representing the portion of the beam line where ions can be accumulated.

c. Neutralization drops instantaneously to zero when the beam is interrupted, increases linearly with time until reaching an equilibrium at some value $\eta_0$, and stays constant afterward:

$$\eta(t) = \begin{cases} t/\tau_c, & t \leq \tau_0 \\ \eta_0, & t > \tau_0 \end{cases} \quad \tau_c = \frac{1}{n_a \sigma_i \beta c}; \quad \tau_0 = \eta_0 \tau_c \qquad (19)$$

where $n_a$ is the atom density and $\sigma_i$ is the ionization cross section of hydrogen.

d. The longitudinal cooling force $F_c$ changes with the additional envelope angle introduced by neutralization as

$$F_c = \frac{F_0}{1 + (\Delta \alpha_i / \alpha_0)^2}, \qquad (20)$$

where $\alpha_0$ and $F_0$ are the rms angle and drag force at optimum focusing.

e. The measured drag rate $F_d$ is the cooling force averaged over the period between interruptions (assuming that the pencil antiproton beam is sensitive mainly to the electron angles in the location of its center)

$$F_d = f_{int} \int_0^{1/f_{int}} F_c(t)\,dt \qquad (21)$$

The model described by Eqs. (18)-(21) was compared with the measurements using the measured value of $F_0 = 73\,\text{MeV/c/hr}$, the rms angle estimated at the beam center $\alpha_0 \approx 0.1\,\text{mrad}$, and fitted parameters $\delta = 0.5$, $\eta_0 = 0.02$. The fit, shown by the dashed line in Figure 17, follows the experimental data reasonably well. In part, it means that ion clearing with only a voltage applied to the BPM plates was already quite effective, ~2%, though still not good enough for this specific task.

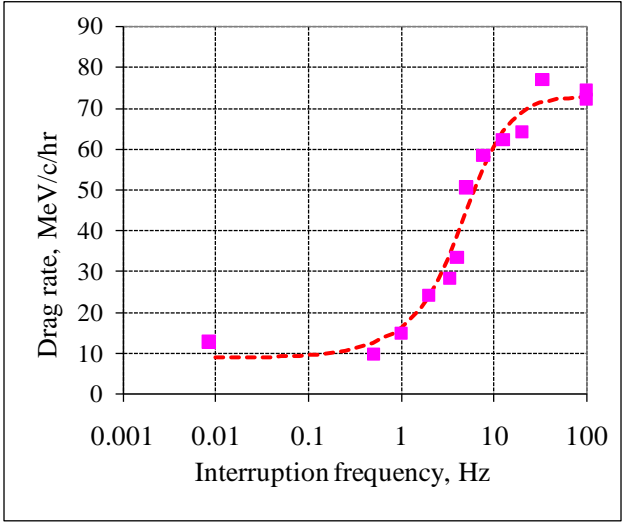

Figure 17. Drag rate as a function of the interruption frequency $f_{int}$ for $I_e$ = 0.3A and separation between beams of 1 mm. January 2, 2011. The interruption pulse duration was 2 µs. Focusing was optimized on axis at 20 Hz. The squares represent the data, and the line is the model.

For operation, the most important result of ion clearing is an increase of the longitudinal cooling rate by a factor of 2 (see Figure 30) with otherwise similar parameters.

### 6.5. *Coherent dipole motion*

A coherent dipole motion of the beam in the cooling section adds to the transverse velocity of individual electrons hence decreases the cooling efficiency. This motion can be caused either by errors in the beam entrance position and angle or by dipole magnetic fields inside the cooling section.

The main sources of slow (longer than minutes) changes in the position of the beam coming toward the cooling section are mechanical drifts of focusing elements and drift of their power supplies. This contribution is suppressed by a software feedback loop, which adjusts currents of two pairs of dipole correctors immediately upstream of the cooling section according to position measurements in two pairs of BPMs inside the section. The measurements are synchronized with the Main Injector cycle and, therefore, are typically made every 2.2 sec.

The original plan also foresaw the use of a faster (up to ~10 Hz) feedback loop to compensate the effect from the MI stray fields, but it was not implemented mainly because the corresponding oscillations were found to contribute insignificantly into the angle budget. This was determined from an analysis of time-dependent readings of the cooling section BPMs [39]. For every time sampling point, deviations of the positions in 11 BPM pairs were fitted to a helical trajectory. The fit was found to be close to the raw signal, so the BPM noise contribution was small. Then, the beam position and angle at the cooling section entrance were reconstructed from each fit. The FFT of the resulting set showed the MI field contribution to be ~30 μrad rms. A similar procedure applied without MI ramps [40] showed that contributions from the Pelletron vibrations (mainly 20, 29.8, and 59.5 Hz lines [41]) and the power grid (60 Hz) accounted together for ~20 μrad rms (1D).

Because the cooling section has good magnetic shielding, external magnetic fields are decreased by a factor of >1000 inside [42] and do not create any measurable perturbations. On the other hand, dipole magnetic fields created by imperfections in the cooling section magnetic field are significant and typically the main component of the total angle. After installation, the transverse magnetic fields were measured and compensated with dipole correctors to the level where the resulting rms angle was estimated to be ~50 μrad [43]. However, beam trajectories measured several months afterwards indicated much larger values. Because studies at the R&D stage proved that the individual 2-m modules are rigid, it was concluded that the reason for the trajectory's drift is mechanical shifts of the modules with respect to one another. Indeed, to produce a helix with a 1D rms angle of 100 μrad, it is enough for one of the supports of a solenoidal module to be shifted by ~0.1 mm, which is modest in comparison with millimeter-range tunnel drifts observed after construction of the Electron cooler's building. Later, there were also hints that the cooling section deforms significantly (in this scale) during fast changes of the cooling section temperature, for example, as result of a cooling water system failure.

These drifts were compensated multiple times during operation of the cooler using the cooling section dipole correctors. Each of ten 2m-long modules of the cooling section is equipped with 8 pairs of 20cm-long printed circuit dipole correctors (so called main correctors), mounted outside of the solenoid winding along the entire length of the module, and 2 pairs of narrow correctors attached to the inner surface of the solenoids at each end of the module (so called end correctors). A simultaneous change of all main correctors' currents in a module by the same amount creates a dipole field roughly equivalent to an inclined solenoid (with respect to its initial position); adjustment of an end corrector has an effect similar to a shift of the module end with respect to the neighboring module. Hence, the right combination of these two types of adjustments should be able to compensate the mechanical drifts. The difficulty is in determining the appropriate values of the corrector currents. Note that the simple alignment of the beam to the centers of the cooling section BPMs with, for example, the end correctors only, does not guarantee the straightness of the trajectory. In fact, practical attempts to make alignment in this manner resulted only in worse cooling.

After testing several procedures, the following one, based on the drag rate measurements (see section 8), was eventually developed and implemented.

1. The antiproton beam trajectory is measured with the cooling section BPMs in order to 're-align' them. Because of the large antiproton momentum and effective magnetic shielding, this trajectory is a straight line in comparison with wiggling of the electron beam. Therefore, deviations from the line reported by BPMs results from the mechanical offsets of the BPMs. These deviations are then subtracted in the software that calculates the beam positions. Thus, once implemented, the measured trajectory should again be a straight line within the measurements errors. Before installation all BPMs were calibrated at a bench in both antiproton (89 kHz, the Recycler revolution frequency) and electron (32 kHz, frequency of the electron beam current modulation) modes [22]. Hence, the subtraction corrects the measurements in the electron mode as well.
2. A pencil antiproton beam is prepared for the dag rate measurements.
3. The electron beam is quickly shifted far from the antiprotons everywhere in the cooling section except for one module (see an example in Figure 18), and the drag rate is measured. The electron trajectory is returned to the standard position to ensure the same initial conditions of the antiprotons in each measurement.
4. All ($x$ or $y$) main correctors of the module are changed by the same value, and step (3) is repeated. During the measurement, the electron beam is kept centered in the BPMs at both ends of the module.
5. After repeating step (4) for several values of the correctors current both in $x$ and $y$, the values corresponding to the largest drag are determined. A typical example of the measurement result is shown in Figure 19.
6. After measuring all modules, the changes of the main correctors providing the best drag rates are applied. Finally, the resulting trajectory is aligned to the centers of all BPMs with the corresponding end correctors.

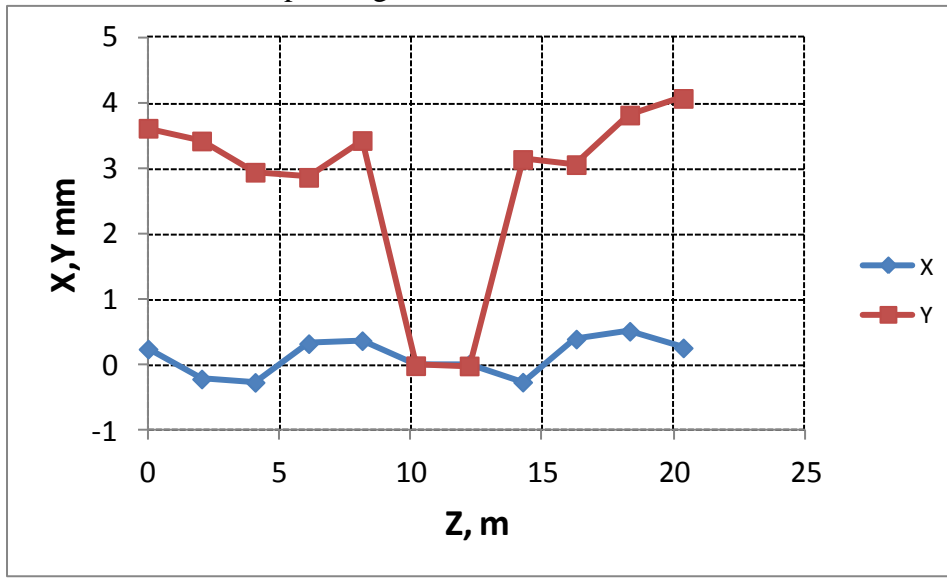

Figure 18. Trajectory in the cooling section when measuring the drag rate in the module positioned between 10 and 12 m. Points show data from BPMs located at the beginning of each module.

Because for a single module the drag rate is low, the statistical noise is high, and the procedure allows correcting only large errors. Also, it is time-consuming (~20 hrs) and was repeated mainly after the complex's shutdowns, when deformations of the cooling section were maximal.

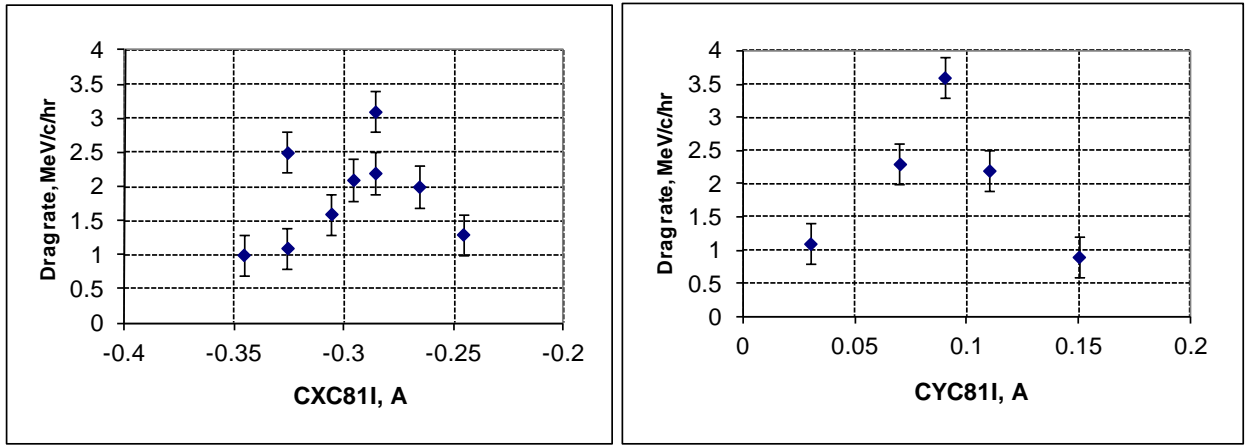

Figure 19. Typical measurements during the cooling section field alignment. First, the drag rate of a single module is measured as a function of changes to all X main correctors in section #8 (left plot). The horizontal axis is the current of one of these correctors. Then, the X-correctors are set to the "optimum" values, and the measurement is repeated with the Y main correctors. In this specific set, the currents of the X correctors were increased by 40mA i.e. CXC81I was changed from -0.326 A to -0.286 A, and the Y correctors currents were lowered by 20mA i.e. CYC81I was changed from 0.11 to 0.09A. The error bars show typical statistical errors of individual measurements.

One can subtract the corrector settings before and after the procedure and feed them into a tracking program. The resulting differential trajectory typically yields a 1D rms angle ~ 100 μrad, indicating the scale of the possible improvement.

### 6.6. *Summary of electron angles*

Table 2 summarizes the contributions from the various effects discussed above to the total electron angle in the cooling section for the best tuning of each component. The values are obtained from averaging the angles over the transverse section of the beam assuming a 2mm radius, and using the angles estimates for the beam current of 0.1 A. Obviously, different ways of averaging are relevant for different scenarios, so the table mainly shows the scale of the effects. For the calculation of the total angle, individual contributions are summed in quadratures.

Table 2. Contributions to the total electron angle in the cooling section. Shown values are 1D, rms.

| Effect | Angle, μrad | Method of evaluation |
|---|---|---|
| Thermal velocities | 57 | Calculated from the cathode temperature |
| Envelope mismatch | ~50 | Resolution of tuning + OptiM simulations |
| Dipole motion (above 0.1 Hz) | ~35 | Spectra of BPMs in the cooling section |
| Dipole motion caused by field imperfections | ~50 | Simulation of electron trajectory in the measured magnetic field |
| Non-linearity of lenses | ~20 | Trajectory response measurements |
| Ion background | < 10 | Cooling measurements |
| **Total** | ~100 | Summed in quadratures |

## 7. Electron energy spread, stability, and matching

The effective energy spread of the electrons in the cooling section is determined by intra-beam processes and by fluctuations of the Pelletron terminal voltage, with the contribution of the longitudinal thermal velocities at the cathode being negligible. A closely related effect, very important for operation, is a slow drift of the electron energy.

### *7.1. Intra-beam processes*

Core of the energy distribution is formed by multiple intra-beam scattering (IBS) and by the dissipation of density micro-fluctuations. For the case relevant for the Fermilab cooler, when the electron transverse temperature is much higher than longitudinal (in the beam frame) and the beam envelope is determined by conservation of the canonical momentum, the resulting rms energy spread was estimated in Ref. [31] to be ~90 eV.

### *7.2. Terminal voltage ripple*

Because the typical cooling time is minutes, the effect of fluctuations of the electron energy on the cooling process caused by the terminal voltage ripple at Hz – range frequencies is heavily averaged and equivalent to an increase of the beam energy spread. The most accurate way to measure these fluctuations was found to be by analyzing the BPM readings in a high-dispersion region [41], [44]. Optical analyses of different frequency components showed measurable energy fluctuations only at frequencies in the 1 – 6 Hz range. Lower frequencies are effectively suppressed by the energy stabilization system, and the higher frequencies are eliminated by the large capacitance of the terminal shell (~300 pF). The typical rms value of the terminal voltage ripple is about 150 V.

The ripple comes primarily from the chain current fluctuations at the chain rotation frequency of 1.8 Hz and its harmonics [44] and depends on the chain condition, and settings of the energy stabilization system. This system uses a Generation Volt Meter (GVM) with a DC to ~10Hz bandwidth as the main tool for measuring the electron energy. In addition, there are capacitive pickups mounted on the Pelletron tank opposite to the terminal shell but their circuitry was found ineffective and they were not used for HV regulation. More details on the performance of the energy regulation system can be found in Ref. [44].

### *7.3. Terminal voltage drift*

Several mechanisms responsible for the energy drift and corresponding solutions are listed in Table 3. In addition to stabilization of the temperatures, two software loops were implemented. One of them adjusts the chain current to eliminate the difference between setting and reading of the terminal voltage.

The second loop modifies the set point of the terminal voltage to keep constant the beam position in a high-dispersion region right after the 180° bend that follows the cooling section. Because the beam position in the low-dispersion (<10 cm) cooling section is stabilized (see section 6.5) and the field in the bend magnet is regulated with NMR sensors, a fixed position implies a constant energy. After 0.1 Hz filtering of the BPM signal, the residual beam motion and electronics noise limit the resolution to about 30 eV. It is enough to keep the energy at the right value. However, failures of the NMR system caused by high radiation in tunnel, tunnel temperature variations, and drifts of the trajectory in the cooling section change the calibration of the loop. The most sensitive (in operation) indication of an energy mismatch was the flattening of the Schottky momentum distribution (Figure 20). Approximately monthly, the calibration of the loop is adjusted by making the momentum distribution as peaky as possible.

Table 3. Factors affecting the energy drift.

| | Sensitivity | Cause | Remedy |
|---|---|---|---|
| Variations of the building temperature | 500 eV/K | Temperature sensitivity of the GVM preamplifier | Temperature of the preamplifier is stabilized to within ±0.5K. Software loop. |
| Variations of the Pelletron temperature | 400 eV/K | Distance between the terminal and GVM depends on temperature | The tank temperature is kept within ±0.2 K. Software loop. |
| Chain current drift or corona current from the terminal | 100 eV/μA | Insufficient gain of the analog feedback loop | Software loop adjusting the chain current. |
| SF6 pressure | ~500 eV/psi | Effect of SF6 permittivity on GVM reading | Software loop. |

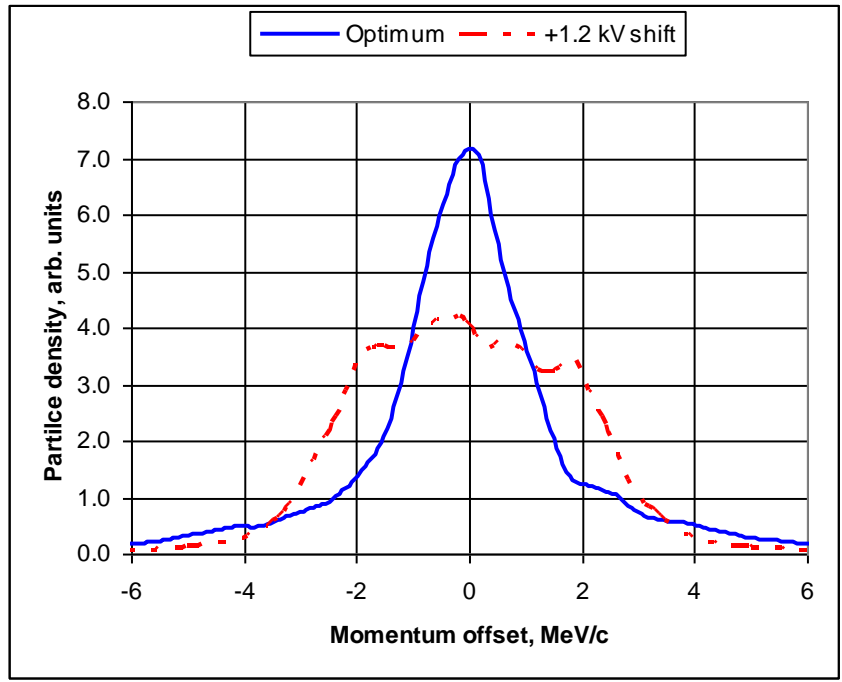

Figure 20. Momentum distribution of the antiproton beam electron cooled at two different electron energies. The solid blue line shows data for optimum energy tuning, and the dash-dot red line represents data for cooling with the electron energy shifted by 1.2 keV. September 10, 2008. $105 \times 10^{10}$ antiprotons, bucket length 5.8 μs; electron beam is on axis. Vertical scale is linear.

### 7.4. Initial energy matching

One of the problems during commissioning of the cooler was the initial matching of the electron and antiproton energies. Because of slow cooling times, the cooling effects are indistinguishable if the energy error is larger than ~0.1%. The absolute energy of antiprotons was known to better than 0.1% from fix-target experiments in the Accumulator and cross-calibrations between machines. The electron energy, determined mainly by the terminal potential, was known initially with much worse accuracy. After assembling the Pelletron, the GVM calibration was verified with a 100 kV external power supply and a calibrated resistive divider. The next step for calibrating the electron energy was to measure the length of a Larmor spiral pitch in the cooling section [45] at the energy close to the nominal. The resulting change of the GVM calibration was 5.2%. The precision of this measurement, determined by the calibration of the Hall probe used for the longitudinal field measurements before installation and by the beam position measurement errors, was estimated to be ~0.2%. It was sufficiently low to observe the first interaction between beams using a specially developed procedure [46]. Antiprotons were smeared over the Recycler momentum aperture (~ ±0.4%), and then electron energy was shifted into the presumed optimum value. Electron cooling created a peak where the antiprotons and electrons energies matched (Figure 21). In the first observation of the cooling force, the electron energy was found to be within 3 keV (~0.07%) of its optimal value.

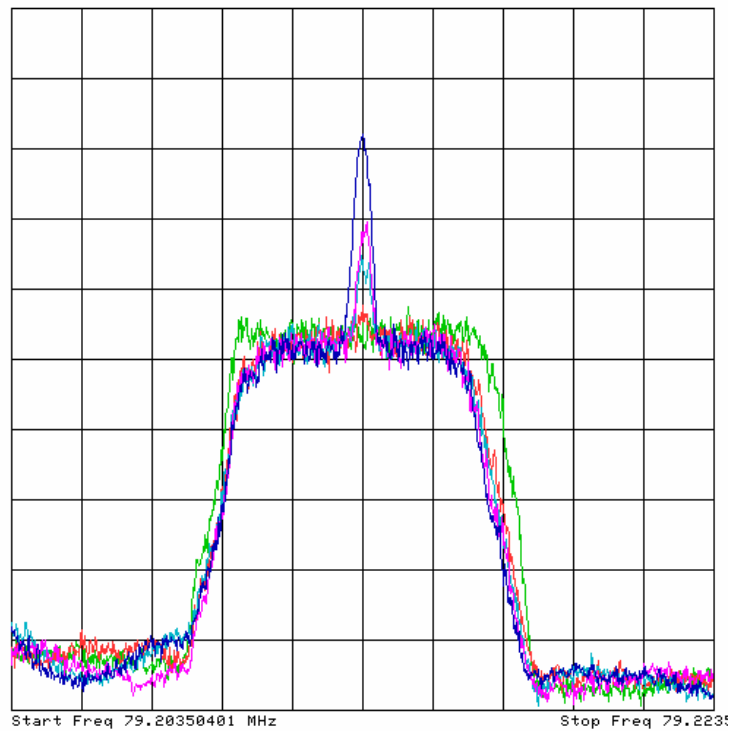

Figure 21. Evolution of the antiproton distribution function after turning on the 0.5A electron beam. The vertical axis represents the longitudinal Schottky power (arb. units, log scale). The horizontal axis is the frequency with the scale, after conversion into momentum units, of 25 MeV/c/div. The antiproton beam intensity was $5\times10^{10}$, and its transverse emittance was 2 μm-rad. Traces were taken 15 minutes apart.

## 8. Cooling force measurements

For operation, the figures of merit to assess cooling are the speed at which the longitudinal and transverse emittances decrease (cooling rates) and the equilibrium emittance values. The procedure of the cooling rate measurements is described in section 9. However, knowledge about details of the electron beam properties can only come from the drag rate measurements discussed below in this section.

## 8.1. Procedure of a drag rate measurement

1. A coasting antiproton beam is scraped down to a low intensity. Empirically, the optimum intensity is determined to be ~$1 \cdot 10^{10}$ particles, large enough for the longitudinal Schottky signal to remain reasonable and low enough to avoid accumulation of ions in the absence of gap in the beam. Longitudinal stochastic cooling is turned off, but the transverse system stays on.
2. With the electron beam "on axis" (i.e. in the standard configuration where the antiproton and electron beam centers coincide), antiprotons are cooled to an equilibrium.
3. If called for, the electron beam state is changed. For example, to study the radial dependence of the cooling force, the electron beam trajectory is shifted parallel to the axis in the cooling section.
4. Immediately after, the electron beam energy is changed by increasing (or decreasing) the terminal voltage set point by $\delta U_t = 0.5 \div 10$ kV (in one step) a.k.a. a "voltage jump".
5. While the antiprotons are dragged to the new equilibrium, their longitudinal distribution is recorded every 15-17 sec, and the average momentum $\overline{p_p}$ and the distribution rms width $\sigma_p$ are calculated.
6. After 2 minutes, the electron energy is returned to its original value, as well as other parameters that might have been changed, and antiprotons are cooled again into the "standard" equilibrium.
7. The drag rate $\dot{\overline{p_p}}$ is calculated from the linear fit of the $\overline{p_p}(t)$ data recorded during $\tau_{meas} = 2$ min with the offset electron energy.

The procedure is illustrated in Figure 22, which shows Schottky spectra recorded at regular intervals after the electron beam voltage jump. Note that in a standard measurement the points are recorded more frequently and for a shorter period of time than in this illustration.

One of the obvious limitations of the procedure is a relatively long measurement time. It was chosen to have a reasonable number of points (5 – 7) for a linear fit, taking into account the significant statistical noise of the mean of the measured distribution. The notion that the drag rate is measured at the momentum offset equal to the initial value of $\overline{\Delta p_{p\,0}} = \frac{M_p}{m_e} e \cdot \delta U_t$ is self-consistent only if the measured rate $\dot{p}$ is low enough so that the shift during the measurement is small in comparison with the initial offset:

$$\dot{p} \cdot \tau_{meas} \ll \overline{\Delta p_{p\,0}} \tag{22}$$

For the most frequently used voltage jump of $\delta U_t = $ 2kV, Eq.(22) limits $\dot{p} \ll$ 100 MeV/c/hr. This inequality was typically satisfied at the early stages of the project but did not remain true as the cooling properties were improved. Because we have not found a better way to measure the cooling force, we still present some data where Eq.(22) is not fulfilled. Therefore, results for small momentum offsets and large drag rates underestimate the value of the drag rate (hence cooling force) that would be measured if a faster measurement could be made.

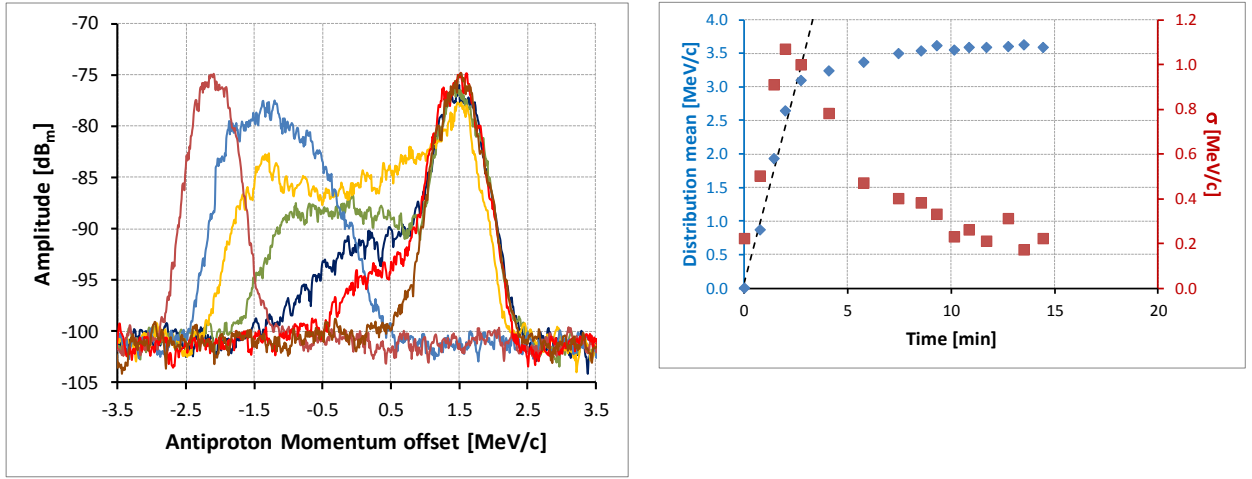

Figure 22. Left - Evolution of the antiproton momentum distribution recorded by a Schottky monitor after a 1.9 keV jump of the electron energy. $I_e$=0.5A with ion clearing at 100 Hz. The time between the first and the last traces is 7 min. January 2, 2011. Right - Corresponding evolution of the mean and rms values of the momentum distribution. The drag rate calculated with a linear fit to the first 4 points is 71 (MeV/c)/hr. (Not a standard measurement).

## 8.2. *Drag rate and the longitudinal cooling force*

The drag rate represents the longitudinal cooling force averaged over all antiprotons. Because these measurements were always made with a pencil beam, one can expand the expression of the force near its value for a central particle. Assuming that the 6D antiproton distribution is symmetrical and identical in horizontal and vertical directions, the instantaneous drag rate can be expressed as

$$\dot{\bar{p}} = \overline{F_{lz}(\Delta p_p, x_p, y_p, p_{px}, p_{py})} \approx F_{lz} + \frac{\partial^2 F_{lz}}{\partial \Delta p_p^2} \cdot \frac{\overline{\Delta p_p^2}}{2} + \frac{\partial^2 F_{lz}}{\partial p_{px}^2} \cdot \overline{p_{px}^2} + \frac{\partial^2 F_{lz}}{\partial x_p^2} \cdot \overline{x_p^2} =$$
$$= F_{lz} + \frac{\partial^2 F_{lz}}{\partial \Delta p_p^2} \cdot \frac{\overline{\Delta p_p^2}}{2} + \left( \frac{\partial^2 F_{lz}}{\partial p_{px}^2} \cdot \frac{p_p^2}{\beta_{CS}} + \frac{\partial^2 F_{lz}}{\partial x_p^2} \cdot \beta_{CS} \right) \varepsilon_{px}$$

(23)

where all functions on the right side are estimated for the central particle $(\overline{\Delta p_p}, 0, 0, 0, 0)$, $p_p = \gamma \beta M_p c$, $\beta_{CS} \approx 30$m is the beta-function in the cooling section, and $\varepsilon_{px}$ is the antiproton transverse rms emittance. Assuming applicability of Eq.(7), both second derivatives over the antiproton momentum have the sign opposite to the cooling force. The measured transverse dependence of the drag rate approximated below by Eq.(26) gives the opposite sign of a spacial derivative as well. Hence, both correction terms in Eq. (23) decrease the drag rate, and the drag rate is always a lower limit of the cooling force value.

To interpret a drag rate as a cooling force experienced by the central particle, the antiproton beam needs to have a small rms momentum spread $\delta p_p = \sqrt{\overline{\Delta p_p^2}}$ and a small transverse emittance. To estimate the contribution related to the finite momentum spread, let's assume the dependence on the momentum offset is the same for the drag rate and the cooling force and calculate the second derivative of a fit to the measurements. In the case of the example shown in Figure 23, this contribution is small (< 10%) if the momentum offset is > 2 MeV/c at the typical case of $\delta p_p < 1$ MeV/c. The relative value of this correction does not change significantly for lower transverse electron velocities and remained small also for later measurements, when the electron beam quality had been optimized.

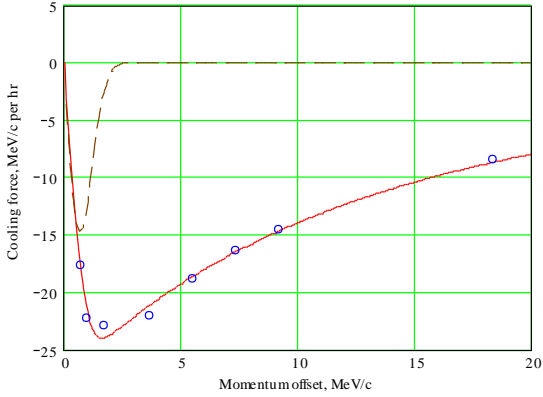

Figure 23. Drag rates measured on February 6, 2006 (blue circles). $N_p = 3.5 \cdot 10^{10}$ antiprotons, transverse emittance (95%, normalized, measured with a Schottky pickup) is $\varepsilon_{p,95\%n} \approx 1.5\pi$ mm·mrad, initial rms momentum spread $\delta p_p = 0.2$ MeV/c, electron beam current $I_e = 0.1$ A. The red solid line represents Eq. (8) at $\delta W_e = 370$ eV, $\theta_e = 0.2$ mrad (discussed in Sections 6 and 7) and $j_e = 0.94$ A/cm$^2$ (from gun simulations), and $L_c = 14$. Contribution of the second derivative term is shown by the dashed brown line (for illustration purposes, it is negated and shown for $\delta p_p = 1$ MeV/c).

The relative contribution of the second correction term in Eq.(23) can be expressed as

$$\frac{1}{F_{lz}} \frac{\partial^2 F_{lz}}{\partial p_{px}^2} \cdot \overline{p_{px}^2} = f_{px\_contr} \cdot \left(\frac{\vartheta_{px}}{\vartheta_t}\right)^2 \quad (24)$$

where $f_{px\_contr}$ is a function of the electron velocities and the antiproton momentum offset, and $\theta_{px}$ is the antiproton 1D angle in the cooling section. For relevant parameters, estimations from Eq.(7) give $f_{px\_contr} = 1.5 \div 2$, and in drag rate measurements the antiproton angles were an order of magnitude lower than the electrons'. Hence, this contribution is negligible.

The last term in Eq. (23) was found to be significantly more important and can be estimated from the dependence of the measured drag rate on the radial position with the following assumptions:
1. The contribution of far collisions can be neglected, and the force is determined by local values of the electron current density $j_e$ and rms angle at the position of the pencil antiproton beam as in Eq.(9), $F_{lz} \prec \dfrac{j_e}{\vartheta_t^2}$.
2. The electron rms angle is determined by its value at the beam center, $\alpha_0$, and the envelope mismatch angle added in quadratures. The latter is linear with radius, and the total angle can be expressed as $\vartheta_t^2 = \alpha_0^2\left(1 + \dfrac{x^2}{b^2}\right)$ (similar to Eq.(20)).

3. The current density distribution is calculated from the gun current density simulations (see Figure 14) and approximated by a parabola

$$j_e = j_0 \begin{cases} 1 - \dfrac{x^2}{a_e^2}, & x \leq a_e \\ 0, & x > a_e \end{cases} \qquad (25)$$

4. The transverse distribution in the antiproton beam is Gaussian.
5. The cooling force can be presented as a product of a component determined by the momentum offset, $F_0(\Delta p_p)$, and the radial dependence:

$$F_{lz}(x) = F_0(\Delta p_p) \cdot \begin{cases} \dfrac{1 - \left(\dfrac{x}{a_e}\right)^2}{1 + \left(\dfrac{x}{b}\right)^2}, & x \leq a_e \\ 0, & x > a_e \end{cases} \qquad (26)$$

6. The drag rate at a radial offset is the cooling force of Eq. (26) integrated over a Gaussian antiproton radial distribution with the rms size $a_p$.

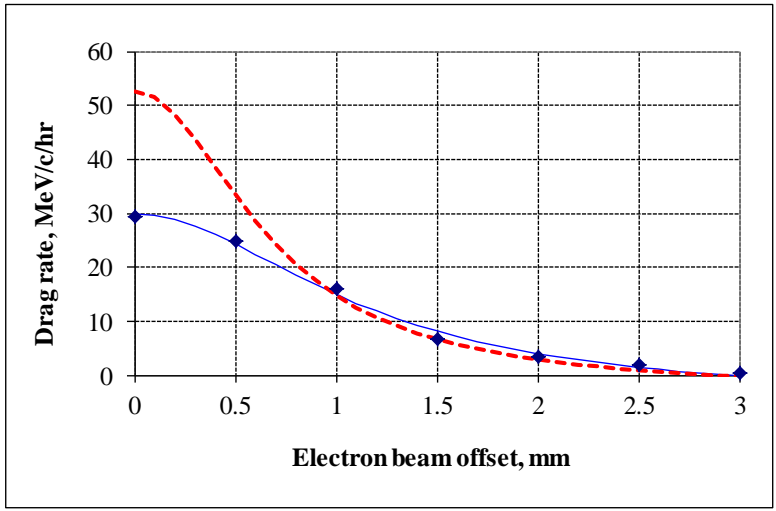

Figure 24. Drag rate as a function of the electron beam offset. The points are results from July 3, 2007 measurements. The voltage jump was 2 kV, $I_e$ = 0.1 A, $N_p$ = 4·10$^{10}$. During the measurement, the emittance measured with a flying wire was $\varepsilon_{p,95\%n}$ ~ 0.5 µm (95%, normalized), which corresponds to $a_p$ ~ 0.5 mm. The red curve is the reconstructed cooling force.

Comparison of the model with the measured drag rates is shown in Figure 24 as the solid curve, which is calculated with $a_e = 2.9$ mm from the gun simulations, $a_p = 0.5$ mm estimated from the flying wire measurements, and the best fitting values of $F_0 = 53$ MeV/c/hr and $b = 0.68$ mm. The cooling force in the center is higher than the measured drag rate by almost a factor of two. Correspondingly, the relative contribution of the last term in Eq. (23)

$$\frac{1}{F_{lz}} \frac{\partial^2 F_{lz}}{\partial x_p^2} \cdot \overline{x_p^2} = -2\overline{x_p^2}\left(\frac{1}{a^2} + \frac{1}{b^2}\right) = -4.56 \cdot 0.5^2 = -1.14. \tag{27}$$

is large, and the expansion of Eq. (23) is not self-consistent.

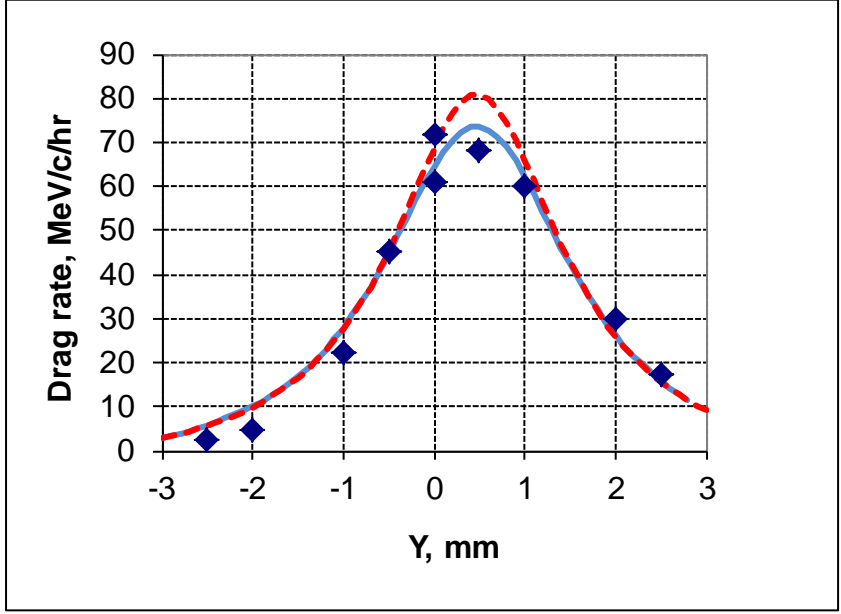

Figure 25. Drag rate as a function of the electron beam offset. The points are results from December 31, 2010 (the same set as in Figure 16b). The voltage jump was 2 kV, $I_e = 0.3$ A, $N_p = 1.3 \cdot 10^{10}$. During the measurement, the emittance measured with a flying wire was $\varepsilon_{p,95\%n} < 0.3$ µm (95%, normalized), and the corresponding value of $a_p$ was estimated to be ~0.25 mm. The blue curve is the best fit to the model described with $a = 4.3$ mm (from simulations), and fitting parameters $F_0 = 80$ MeV/c/hr and $b = 1.2$ mm. Also, to better fit the data the curve is shifted by 0.47 mm in comparison with the position reported by the BPMs. The red curve shows the reconstructed cooling force.

Due to the difficulty to accurately measure and control the transverse emittance, this effect led to a large scatter in the measured drag rates. Eventually to decrease this scatter, the transverse stochastic cooling system was left on during the measurements; the antiproton beam was scraped down to the limit at which a reasonable resolution of the Schottky detector remained, $N_p \sim 1 \cdot 10^{10}$; and between the measurements the electron beam was set to the state of strongest cooling. However, equally important was the decrease of the electron angles spread across the beam as it was discussed in section 6. Consequently, the reproducibility of the results improved. Meanwhile the antiproton beam transverse emittance measured with the flying wire decreased to $< 0.3$ µm (the emittance measurements were unreliable below this level).

As an example, Figure 25 shows a comparison between the drag rates thus measured and the reconstructed cooling force. In this case the drag rates were close to the longitudinal cooling force, and the estimation from Eq.(27) gives a relative contribution of 8%.

### 8.3. Drag rates and electron beam parameters

With all the difficulties associated with the drag rate measurements, they still provided valuable estimations of the cooling force and, consequently, of the electron beam angles at the location of the antiproton beam. Figure 26 shows the dependence of the drag rate on the beam current recorded over the years. All measurements were made with the same voltage jump, $\delta U_t = $ 2kV, and on axis. Each curve was recorded at a fixed set of focusing parameters. If the electron angles stay unchanged, the cooling force should change proportionally to the local current density (shown by the dashed curve). Significant deviations of the drag rates from this trend were the most influential reason for measuring the transverse distributions of the cooling force (as in Figure 16a) and eventually developing the ion clearing procedure described in section 6.4. The results with ion clearing are presented in Figure 26 by the curve labeled 1/2/2011. Note that most of the data points of this set do not satisfy Eq.(22), and the flatness of the curve at larger currents can easily be explained by the inadequacy of the procedure. We were not able to explain and correct the dramatic deterioration of cooling around February of 2011 and the corresponding decrease of the drag rate (curve 3/7/2011), partly because of lack of resources toward the end of the Tevatron operation.

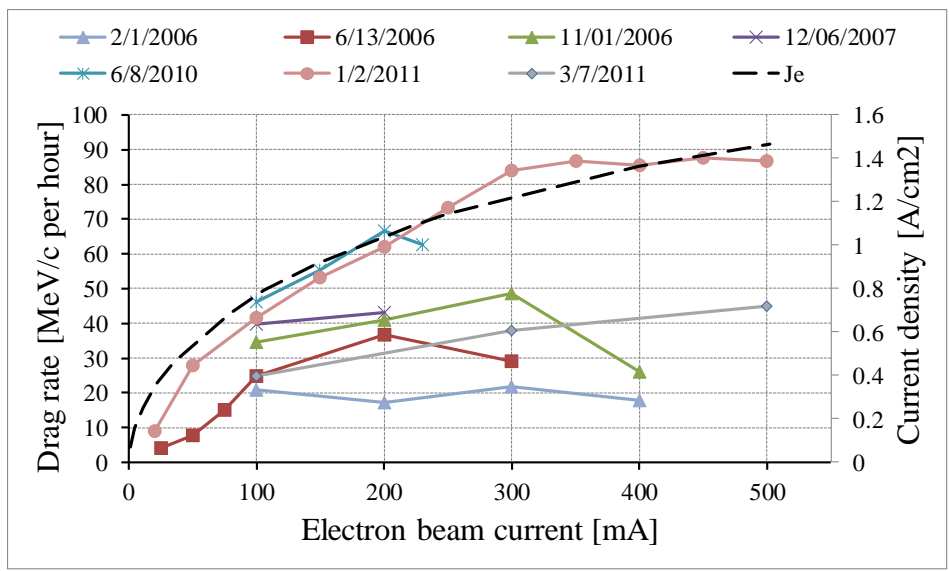

Figure 26. Drag rate as a function of the beam current measured on axis at various dates with a 2 kV voltage jump. The current density calculated at the beam center (dashed curve) is shown for comparison.

Measurements of the drag rate as a function of the voltage jump, interpreted as the cooling force vs momentum offset, give some information about the absolute value of the electron angles in the cooling section. An example of a set of measurements is shown in Figure 27. The data were fitted to Eq. (1) with the Coulomb logarithm under the integral (curve 1) and Gaussian electron velocity distributions. For these measurements, the estimated effect of the final size of the antiproton beam, discussed in the previous section, was small, ≤ 10%, and was not taken into account.

The only fitting parameter was the rms spread of the electron transverse velocities, $\sigma_{et}$, while the longitudinal velocity spread is assumed to be determined by the measured voltage ripple, and the current density is estimated from the gun simulations and magnetic measurements. The first data point i.e. the data point with the smallest momentum offset was excluded from the fit because it was far from satisfying the inequality of Eq. (22). For comparison, curve 2 shows a calculation (for the same $\sigma_{et}$ i.e. no free parameters) using the simplified equation, Eq. (8), with the Coulomb logarithm calculated for each momentum offset according to Eq. (2), (3). Curve 3 is the same as curve 2 but with a constant value of the logarithm estimated at the momentum offset of 10 MeV/c. Taking into account the level of agreement between measurements and the model, the simplified formula gives a good approximation. The value of $\sigma_{et}$ found in the fit, 0.13 mrad, is close to what has been found from other measurements in Table 2, 0.1 mrad.

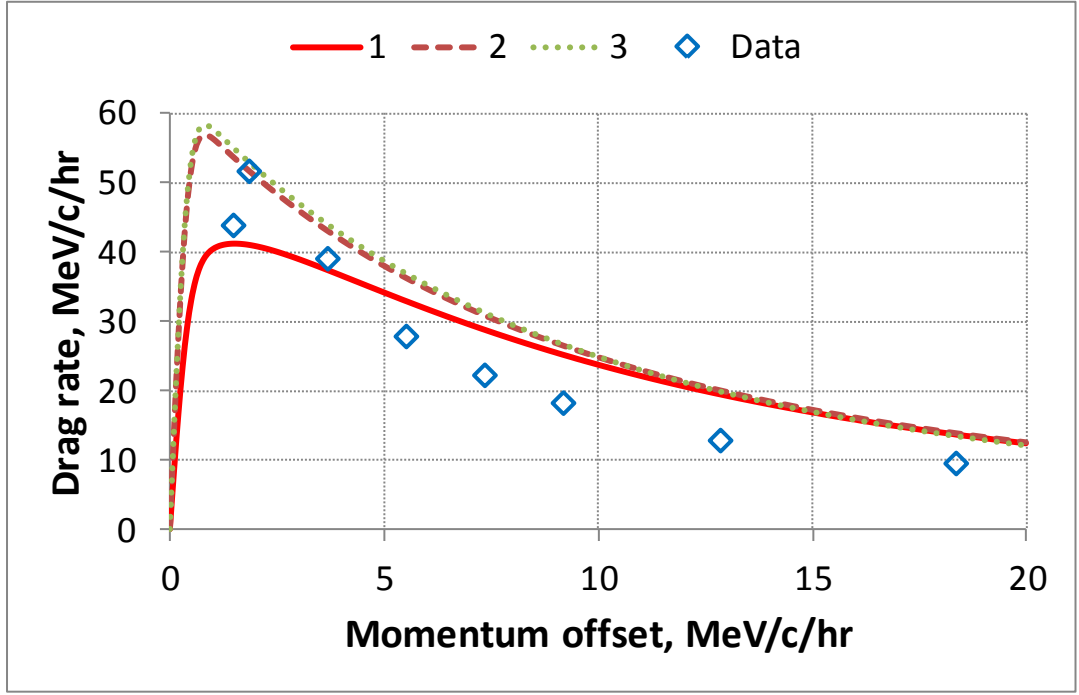

Figure 27. Dependence of the drag rate on the antiproton momentum offset. The diamonds represent data measured on January 4, 2011 with $I_e = 0.1$A "on axis". See the text for the curves description.

A set of measurements was made with the electron beam injected into the cooling section at an angle. In this case, the trajectory of the beam centroid is a helix, and the rms electron angles are increased. As a result, the drag rates decrease (Figure 28), and the fitted value of the angle increases in reasonable agreement with the sum in quadrature of the initial angle $\sigma_{et}(0)$, and the angle introduced by the trajectory perturbation (Table 4).

At the same time, the data points systematically deviate from the fitting curve by more than the statistical measurement errors (which were typically 1-3 MeV/c/hr). It can be corrected by allowing one more fitting parameter. For example, the fit becomes almost perfect if the constant in front of the integral in Eq. (1) is artificially decreased by a factor of 2.5 and $\sigma_{et}$ is set to 0.07 mrad. One may speculate that it is due to a large variation of the value of dipole electron angles between modules in the cooling section, thus only a portion of the section works effectively. However, we did not see such dramatic effects experimentally. More likely reasons for the disagreement are the simplifications made to the model itself.

One of them is the neglect of the longitudinal 105 G magnetic field. One would expect the field to modify the dynamics of binary collisions with large impact parameters (above the electron Larmor radius of ~0.2 mm), but we have not found a way to estimate this effect quantitatively. An indirect indication of the importance of the collisions with large impact parameters was seen in measurements with helix – like trajectories reported in Ref. [47], where the drag rates grew when decreasing the helix pitch at a constant helical angle.

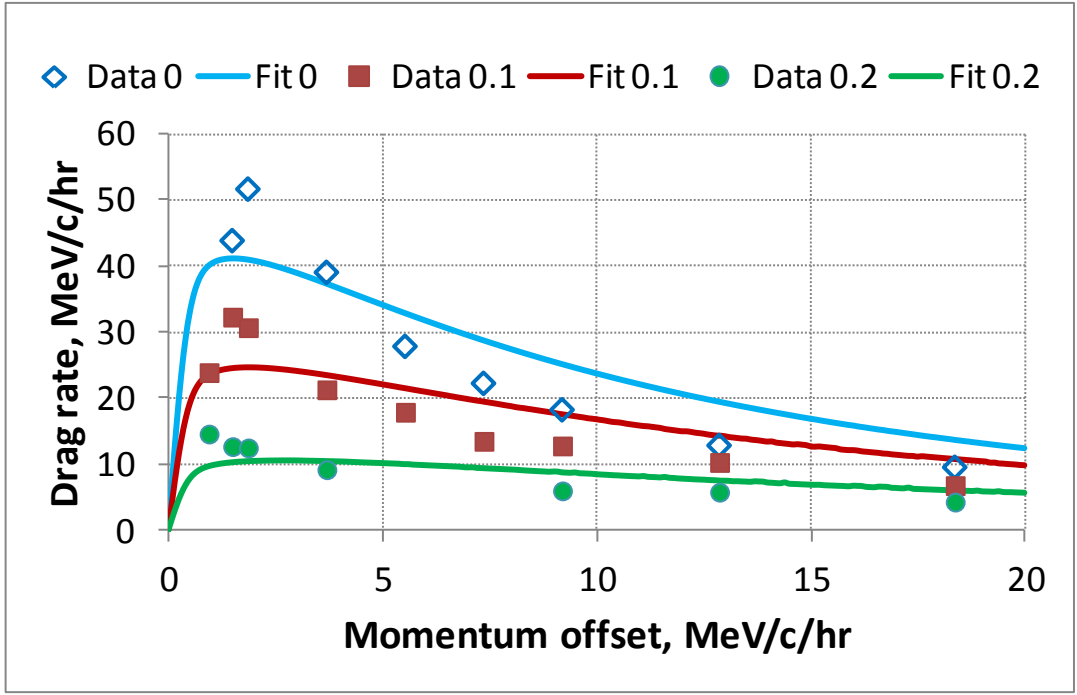

Figure 28. Data and fits with Eq.(1) for a beam with helical trajectories. The numbers in the legend indicate the 1D values of the angle of the beam centroid in the cooling section in milliradian (verified with BPM measurements). The fitted total angles are shown in Table 4.

Table 4. Values of the rms angle, in mrad, from fitting the data in Figure 28 with Eq.(1). The last column compares the fitted angles with the simplest prediction.

| Helical angle (1D), $\alpha_h$ | Fitted rms angle, $\sigma_{et}$ | $\sqrt{\sigma_{et}(0)^2 + \alpha_h^2}$ |
|---|---|---|
| 0 | 0.134 | 0.134 |
| 0.1 | 0.180 | 0.167 |
| 0.2 | 0.285 | 0.241 |

Another possible cause for the disagreement is the assumption that the transverse inhomogeneity of the electron beam affects identically the force at all momentum offsets (Eq.(26)).

From these comparisons, we conclude that the model is capable of predicting the drag rates for weakly-magnetized electron cooling to within ~50%. Using the simplified formula of Eq. (8) makes analytical estimations and fitting much easier and without significant sacrifices in accuracy.

## 9. Electron cooling rates

### 9.1. Definition and procedure

While the drag rate measurements were the instrument to estimate and improve the electron beam properties, the characteristics important for operation were the cooling rates [48]. For example, for longitudinal cooling, one can compare the time derivative of the rms momentum $\delta p_p$ with the cooling system on and off. In the model with uncoupled degrees of freedom and constant diffusion, the derivative depends on the diffusion coefficient $D_p$, the longitudinal cooling force $F_{lz}$, and the antiproton distribution function $f_p$:

$$\frac{d}{dt}\left(\delta p_p^2\right) = \frac{d}{dt}\left[\int p^2 f_p dp - \overline{p}^2\right] = \int p^2 \frac{\partial f_p}{\partial t} dp - 2\overline{p}\cdot\overline{F_{lz}} \tag{28}$$

Assuming symmetry of the distribution and the cooling force, the average force becomes zero, $\overline{F_{lz}} = 0$. Then, using the Fokker-Planck equation with a constant diffusion coefficient, Eq.(23) can be written as

$$\frac{d}{dt}\left(\delta p_p^2\right) = \int p^2 \left[-\frac{\partial}{\partial p}\left(F_{lz} f_p\right) + \frac{\partial}{\partial p}\left(D_p \cdot \frac{\partial f_p}{\partial p}\right)\right] dp$$

$$= 2\int p\left[F_{lz} f_p - D_p \cdot \frac{\partial f_p}{\partial p}\right] dp = 2\int F_{lz} f_p p\, dp + 2D_p \tag{29}$$

Now, we define the cooling rate $R_z$ as the rate at which the rms momentum spread would be decreasing in the absence of any diffusion

$$R_z \equiv \frac{1}{\delta p_p} \int F_{lz} f_p p\, dp \tag{30}$$

If the value of $\delta p_p$ doesn't significantly change during the time of measurements, the cooling rate is close to the difference between its time derivatives with and without cooling

$$R_z = \delta \dot{p}_p\big|_{F_{lz}} - \delta \dot{p}_p\big|_0 \tag{31}$$

The logic is similar for the transverse cooling rates defined through differences of time derivatives of corresponding emittances. Figure 29 gives a typical example of such a measurement.

In order to compare cooling performances for different electron beam tuning, such measurements were done at similar antiproton beam conditions and followed a standard procedure:
1. The antiproton beam, confined by rectangular RF barriers, is first cooled with the stochastic cooling system only. This makes the antiprotons distribution close to Gaussian.
2. The bunch length is adjusted to bring the initial rms momentum spread to ~3.5 MeV/c.
3. The stochastic cooling system is turned off and the antiproton beam is let diffuse for 15 minutes.
4. The electron beam is turned on and cools the antiprotons for 15 minutes.
5. Then, the cooling rate is obtained from calculating the difference between the time derivatives of the momentum spread (or transverse emittances) before and after turning on the electron beam. For the transverse direction, both diffusion and cooling are fitted with straight lines. For the longitudinal direction, the cooling data are fitted to an exponential decay curve and the time derivative is calculated at the time when the electron beam is turned on. On Figure 29, the transverse cooling rate (averaged over both directions) for the Schottky data is -2.4 π mm·mrad/hr and for the flying wire data -5.6 π mm·mrad/hr. For the momentum spread, the measured cooling rate is -7.3 MeV/c per hour.

Note that the cooling rate measurements were almost always performed with the antiproton beam contained in the barrier bucket, while Eq. (28) - (31) assume no synchrotron motion. In the simplest model with the infinite height of barriers, flat bucket bottom, and equal average electron and antiproton velocities, the results are exactly the same. For typical conditions (as for Figure 29), deviations from this ideal picture were not significant: the length of the bucket was much longer that the depth of particle penetration into barriers; the typical momentum offset was well above than deformations of the bottom of the potential well and than the velocities misalignment. Corresponding estimations showed the cumulative effect of presence of the barriers is small in comparison with the measurements scatter.

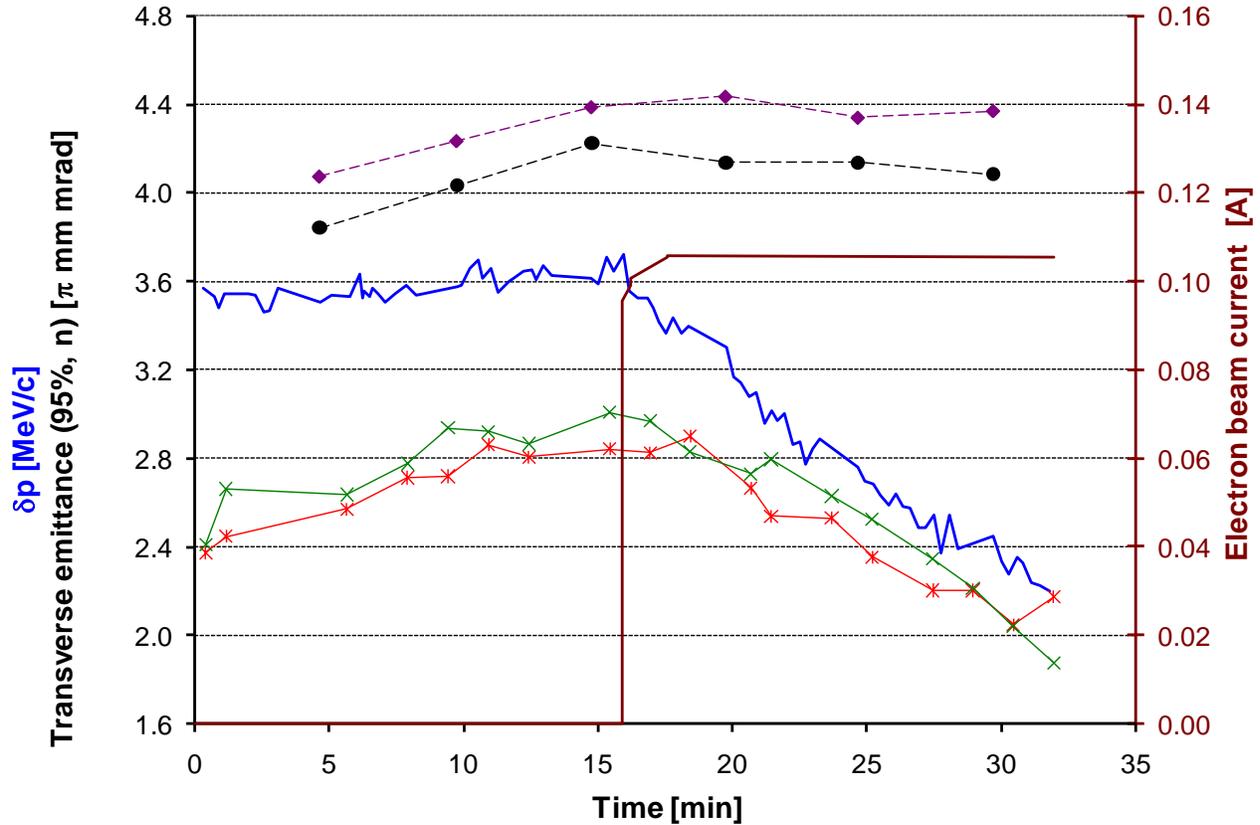

Figure 29. Evolution of $\delta p_p$ (blue line) and of the transverse emittances (95%, normalized) during a cooling rate measurement. The brown line indicates when the electron beam was turned on and set to 100 mA. $N_p = 36 \times 10^{10}$, bunch length = 5.4 µs. For the transverse emittances, both measurements obtained from a Schottky pickup (violet diamonds for X and black circles for Y) and from flying wires (green crosses for X, and red stars for Y) are displayed.

The relatively large difference between the transverse cooling rate obtained from the Schottky data and the flying wire data is reproducible. It is attributed partly to a possible mis- calibration of one of the detectors and partly to the respective sensitivity of these devices in combination with the cooling properties of the electron beam. The emittance measured by flying wires is calculated from a Gaussian fit, hence essentially ignoring the far tails. On the other hand, the Schottky pickup is sensitive to all particles, and, therefore, the weight of particles with large betatron oscillations in the Schottky detector measurement is larger than for the flying wire's. Since electron cooling is less efficient for antiprotons in the transverse tails, the lower cooling rate measured with the Schottky detector seems natural.

## 9.1. Measured longitudinal cooling rates

The notion of the cooling rate has proved to be a useful tool in assessing the performance of the electron cooler. Because the most important cooling component was longitudinal, most of attention was paid to longitudinal rate measurements. Figure 30 summarizes most of the electron cooling rate measurements made between 2006 and 2010. Over that time, the cooling rate for a given transverse emittance significantly increased due to several key improvements to the electron beam quality highlighted on the plot (and discussed in detail in section 6). The arrows indicate the observed cooling rate increase resulting from each of the beam optimization steps. Note for instance that the maximum cooling rate measured for a ~3 µm transverse emittance antiproton beam (from flying wire data) is almost twice the cooling rate obtained from the measurement shown on Figure 29, -13.0 MeV/c per hour vs. -7.3 MeV/c per hour.

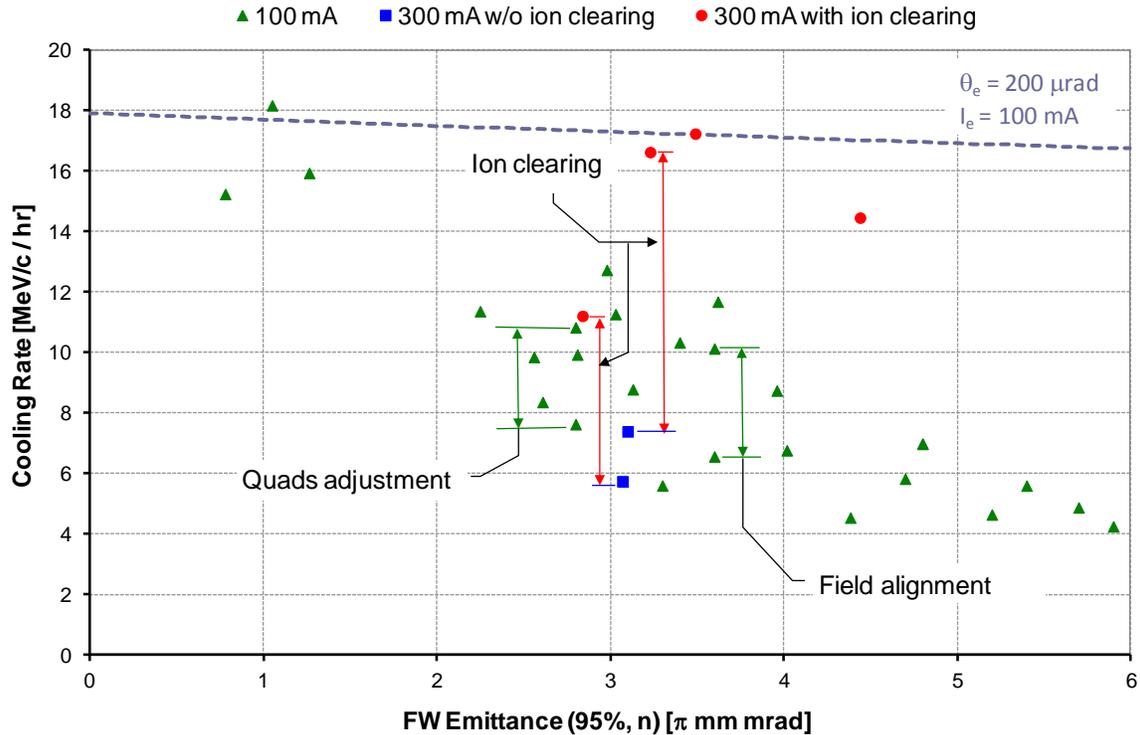

Figure 30. Longitudinal cooling rate (negated) in 2006-2010. The arrows connect data points measured successively over the same day to show the potential cooling rate increase for several improvements brought to the electron beam tuning. The initial rms momentum spread is 3-4 MeV/c. The dashed line represents the expected cooling rate calculated with Eq. (11).

Also visible on the plot is a fairly strong dependence of the cooling rate on the transverse emittance of the antiproton beam. This behavior is not reproduced by the model described in section 4. The dashed line in Figure 30 shows Eq. (11) expressed in the lab frame. It is computed with the following parameters:

- Antiproton initial momentum spread $\delta p_p$ = 3.6 MeV/c, typical for the cooling rate measurements;
- Electron current density $j_e = 0.94$ A/cm$^2$ at the beam center. It was obtained from simulations of the electron gun for an electron beam current $I_e = 0.1$ A;
- Energy spread $\delta W_e$ =175 eV, obtained from high voltage stability measurements;

- Coulomb logarithm $L_c = 15$, calculated for the rms values;
- 1D rms value of the electron angles $\theta_e = 200$ μrad, obtained from fitting cooling rates at low transverse emittance.

Obviously, the measured cooling rate drops with the increase of the antiproton beam emittance much faster than in the model.

The main drawback of the model is the assumption about the transverse homogeneity of the electron beam properties. A strong radial dependence of both the electron current density and angles dramatically decreases the efficiency of cooling for antiprotons traveling through the cooling section at large offsets. For example, in Figure 24 the area of effective electron cooling is limited to <1 mm offset, while for a typical emittance $\varepsilon_{p,95\%n} \approx 4$ μm, the rms antiproton beam radius is 1.4 mm. Therefore, a large fraction of antiprotons interacts with electrons very inefficiently.

A more accurate comparison between drag rate and cooling rate measurement results can be made by including explicitly into Eq. (30) the radial dependence of the cooling force reconstructed from the drag rate measurements as in Eq.(26). Then, for the integration, we assume that the longitudinal and transverse distributions are uncoupled and Gaussian. While generally speaking it is incorrect (for instance in the case of strong electron cooling, see Fig.3 in [49]), this approximation is likely appropriate for the cooling rate measurements.

Results of such calculation with parameters derived from data presented in Figure 24 are shown in Figure 31 (dash-dot pink curve) and compared with the subset of data from Figure 30 measured with similar electron beam conditions.

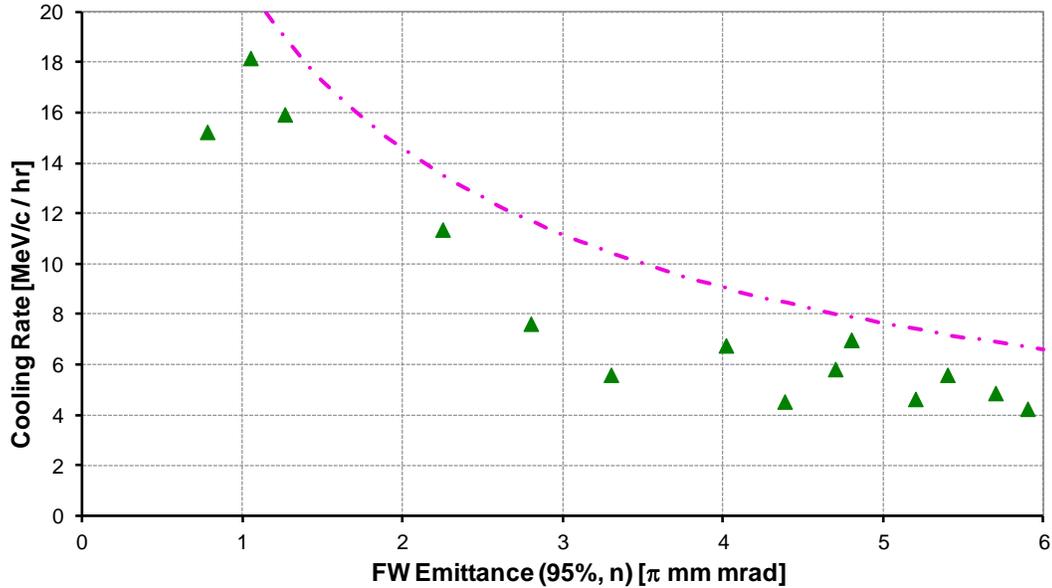

Figure 31. Longitudinal cooling rate (negated) as a function of the antiproton emittance for $I_e = 0.1$ A. The pink dash-dot curve is the calculation with the model described in the text and parameters reconstructed from the drag rate measurements from July 3, 2007. The green triangles are a subset of the data shown on Figure 30, for which the electron beam characteristics are similar (data of February – June 2007).

While this approach still slightly overestimates the cooling rate, it catches well its dependence on the antiprotons transverse emittance. Therefore, one can conclude that the drag rate and cooling rate measurement procedures give consistent descriptions of the cooling properties.

# 10. Electron cooling in the Recycler operation

The Recycler had mainly two functions: store antiprotons efficiently, i.e. with beam loss as small as possible during accumulation, and prepare the beam for extraction to the Tevatron. However, quickly after implementing electron cooling into operation, it was found that fast cooling may result in a significant degradation of the antiproton life time. Correspondingly, the procedures were optimized to balance the cooling speed and antiproton loss.

## *10.1. Electron cooling and the antiprotons life time*

The scale for the cooler's effect on the life time is set by the vacuum-determined limit: a low density antiproton beam cooled with stochastic cooling system had the life time of ~1000 hrs. If this value was staying constant during a typical storage time of ~15 hours, the total loss would be ~0.75% of the maximum amount (assuming a linear rise of the intensity with time). However, in operation the life time averaged over a storing cycle varied from ~500 down to ~100 hours.

Degradation of the life time after application of electron cooling has been observed in low-energy coolers as well (see, for example, Ref. [50]). However, the conditions were dramatically different from the Recycler's and not easily scalable. Also, there is no commonly accepted explanation for the phenomenon.

In the Recycler, attempts to separate the contributions to the life time deterioration from direct interaction between the beams and from effects related to changes in the antiproton distributions caused by electron cooling were inconclusive. One of the difficulties is the large scatter of the life time data measured over the years.

The scatter has several sources. One of them is a "long memory" of the beam: the life time might be different well beyond measurement errors for similar beam parameters (intensity, bunch length, 95% longitudinal and transverse emittances measured with a Schottky detector) hours after an adjustment, injection, or beam manipulation has taken place. The likely reason is a slow rate of changes in the far beam tails, which are not well measured, at large beam intensities.

Another source of the scatter was the tuning state of the stochastic cooling system that was not easy to quantify in operation. The life time was critically dependent on the settings of the stochastic cooling system, even though for operational-size antiproton stacks its cooling rates were by more than an order of magnitude lower than that of the electron cooling. On one occasion when the stochastic cooling system was not operational (and turned off), the life time of the electron-only cooled beam was less than 50 hours even though the electron beam parameters were changed multiple times in an attempt to counteract the life time degradation. The latter is qualitatively understandable, because the transverse tail particles approaching the aperture interact negligibly with the electron beam but strongly with the stochastic cooling system.

On a longer time scale, an additional contributor to the life time was the the Recycler vacuum pressure. A small leak or argon accumulation near transfer lines were changing the life time up to a factor of two.

With all these uncertainties, the main observation was that in general the life time was worse for higher antiproton densities. One of the attempts to quantify this observation is presented in Figure 32. It shows data when there were no transfers or external parameter adjustments for a period of more than an hour and for which the beam emittances had reached an equilibrium. One of the indications of being at an equilibrium is a nearly constant ratio of the average longitudinal and transverse velocities in the beam frame, calculated from the emittances and the bunch length (Figure 32c). The data were selected for a period of a month, when there were no indications of significant variations of the Recycler vacuum or stochastic cooling system tuning. From the many possible ways of plotting the lifetime data as a function of various electron and antiproton beam parameters, the measurements scatter appears to be minimal when plotted against the peak beam current (Figure 32a). For comparison, on Figure 32b the same data are plotted as a function of the calculated space charge tune shift for the central particles (assuming a Gaussian transverse distribution and a constant linear density). Note that for this data set the transverse position of the electron beam (see Chapter 10.2) could be anywhere from 1.6 to 3 mm.

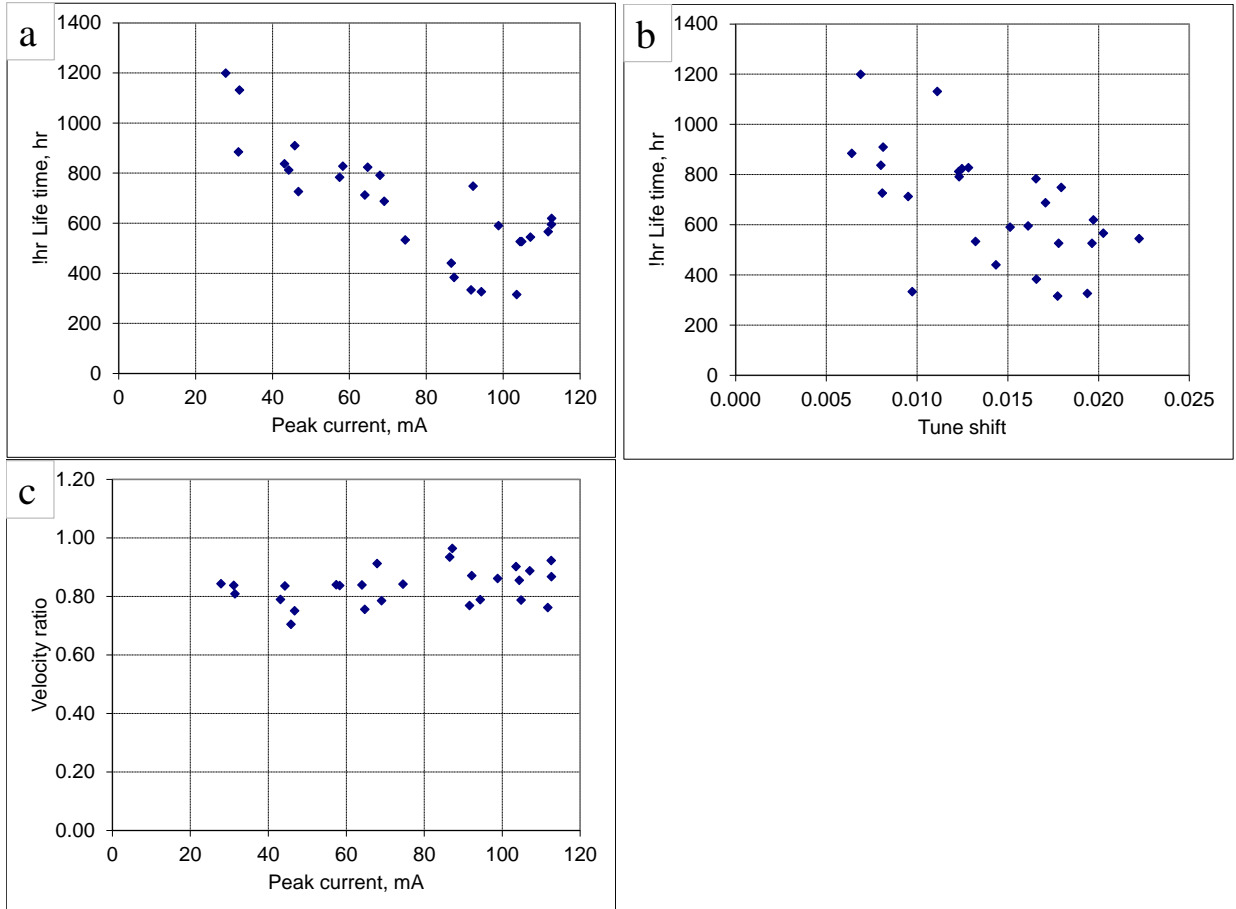

Figure 32. Measurement of equilibrium antiproton parameters. December 2, 2008 – January 7, 2009. The electron beam current was 0.1A for all points.

Once, when a total of ~20 hours was available for measurements, the dependence of the equilibrium life time on the peak current was measured more rigorously (Figure 33). The beam was kept between two rectangular RF barriers, with the initial bunch occupying almost the entire ring circumference. Then every ~ 3 hours, the bunch length was reduced. During the last hour of each measurement, the average life time was constant. Over the duration of the measurements, the 0.1A electron beam was at a fixed 2 mm offset. The transverse emittance was larger for the shorter bunches and consequently the calculated space charge tune shift was not linear with the peak current. Nevertheless, the trends shown on Figures 32 and 33 are similar: the life time decreases with the peak current.

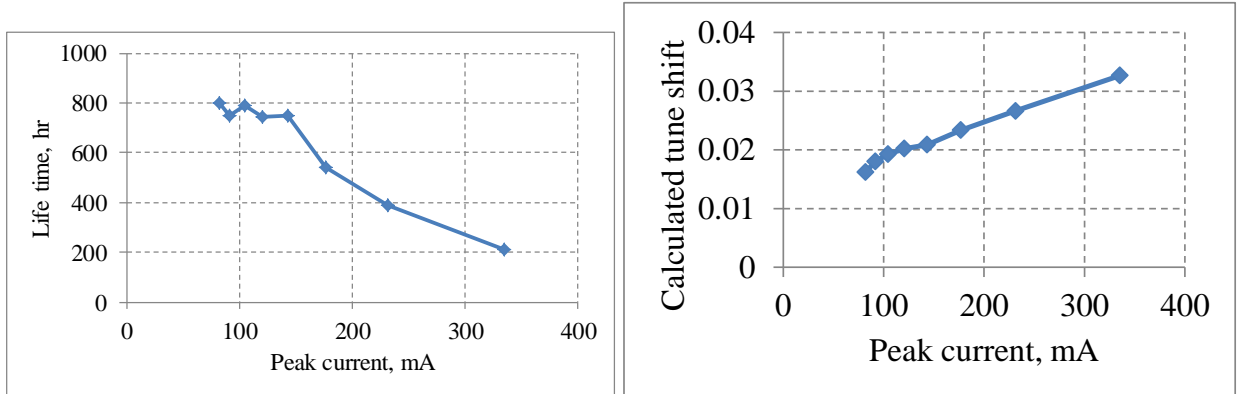

Figure 33. Measurement of the beam life time while reducing the bunch length of the antiproton beam. Left: dependence of the life time on the beam peak current. Right: the space-charge tune shift calculated for central particles. December 26-27, 2010. The number of antiprotons was (446 – 430) $10^{10}$.

It is very likely that the main effect causing deterioration of the life time when electron cooling is applied was indirect, i.e. related more to the denser antiproton distribution created by the electron cooling process and associated changes in the diffusion rather than to the presence of the electron beam itself. One of possibilities is a diffusion induced by envelope instability of the type suggested in Ref. [51]. Note that in the measurements presented in Figure 32 and Figure 33 the information about the transverse distribution was limited to the 95% emittance from the Schottky monitors The core could be denser than for a Gaussian distribution, which was assumed for the tune shift estimations.

An additional increase of losses might be related to the correlation between longitudinal and transverse tails in an electron-cooled antiproton beam. Because electron cooling is ineffective for the transverse tails, these particles are not cooled efficiently longitudinally either. As a result, the transverse and longitudinal tails are correlated (Figure 34 adapted from [49]). Such distribution may increase the losses in high-dispersion areas.

One can also speculate that with the significant impact that the stochastic cooling system has on the antiproton life time, modifications of the antiproton distribution by electron cooling may affect the efficiency of stochastic cooling.

On the other hand, it is likely that the electron beam does add some diffusion. A possible indication was a dramatic, from tens of hours to seconds, shortening of the life time of protons circulating in the Recycler (typically, for machine tuning) after turning on the electron beam. It did not depend on the proton beam intensity, was not sensitive to the working point (tunes), but was affected by the electron beam parameters. Analyses of the life time with respect to the electron beam position and trajectory [52] showed that electron beam current fluctuations at the betatron frequency and, correspondingly, resonant kicks of the electrons' electromagnetic field on the protons may lead to the additional diffusion. The same effect would exist for the antiprotons but being decreased by $2\gamma^2=180$ times because of different propagation direction. Still, it would shrink the antiproton life time down to hours. Nevertheless, turning on the tuned electron beam also brings cooling with much shorter damping times of tens of minutes making this heating negligible for the core particles.

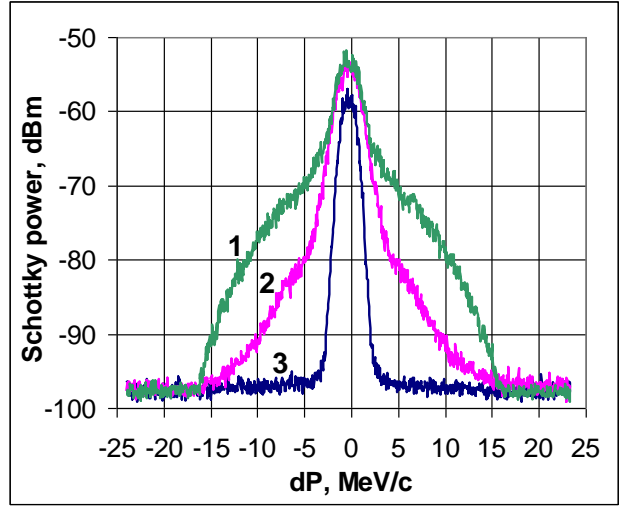

Figure 34. Evolution of the longitudinal 1.75 GHz Schottky profile of a deeply electron-cooled antiproton beam during a vertical scrape. For curves 1, 2, and 3, the number of antiprotons, in units of $10^{10}$, is 26, 20, and 4, respectively. The offset of the vertical scraper, re-calculated into an equivalent cooling section position through the ratio of beta-functions between two locations, is 5.9, 2.1, and 1.7 mm, correspondingly. Before the scrape, antiprotons were cooled with a 0.1 A electron beam for ~ 40 min. The 95%, normalized emittance measured with a flying right before the scrape was 0.7μm (95%, n). In the absence of tail correlation, the longitudinal distribution would not change with removing the transverse tails.

One of the attempts to untangle possible heating and damping terms was to shift the electron energy to suppress cooling (the antiproton energy is essentially fixed by usage of permanent magnets in the Recycler). Unfortunately, even at the offset of ~15 keV, the maximum achievable without major re-tuning of the electron beam line, the main effect was a fast increase of the longitudinal emittance due to the drag force and loss of particles outside of the momentum aperture.

In another set of experiments, the cooling rate was artificially decreased by injecting the electron beam into the cooling section at an angle and thus increasing the effective transverse electron angles. The life time was actually improving, not deteriorating. It seems that the resulting less peaky antiproton Schottky distributions and, correspondingly, lower self – heating was more important than the expected shift of the balance between cooling and possible diffusion from the electron beam. For several months, cooling with the increased angles (with a short wave length similar to [47] to decrease cooling primarily for cold particles) was used in operation, but then was dropped because the modest gain in antiproton survival did not justify an increased complexity of the procedures.

### 10.2. *Electron cooling with a beam offset*

Traditionally, electron cooling is used with the heavy-particle and the electron beams overlapping concentrically, i.e. with the central orbit of the antiprotons in the cooling section and the electron beam centroid trajectory coinciding. This configuration yields the maximum cooling rate but, in the Recycler cooler, is most detrimental for the life time as well.

The practical solution to alleviate this issue, proposed in [53], was to displace the electron beam trajectory parallel with respect to the antiproton beam orbit (a horizontal or vertical shift).

To first approximation, the center of the electron beam is where the current density is the highest and the beam angles the smallest, hence the cooling force the largest. Therefore, when the beam is offset, the majority of the antiprotons do not experience the maximum cooling force. On the other hand, the tails of the antiprotons transverse distribution typically extend further than the electron beam transverse size. Thus, with the beam offset, particles with large betatron oscillations are being cooled more effectively. This procedure can be regarded as 'painting' and, for the Recycler cooler, is almost equivalent to the 'hollow beam' concept for low energy coolers [54] (though such beam offset would likely not work at low energy because of the significance of space charge – related variation of the electron energy across the beam and an increased contribution of diffusion caused by the electron beam current noise). An illustrative cartoon of cooling with an offset is shown in Figure 35.

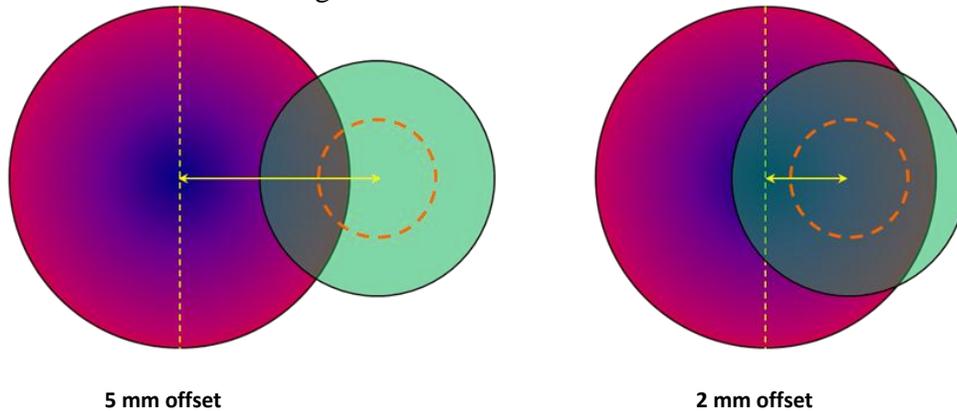

**5 mm offset**  **2 mm offset**

Figure 35. Illustration (to scale) of the cooling rate adjustment by a horizontal shift. For this schematic the antiproton beam size is 4.2 mm ($\pm 2\sigma$) and the color scheme gradient represents the Gaussian particles distribution. The electron beam diameter is 3 mm. The 'area of good cooling' represents the transverse section of the electron beam where the current density and the beam angle are the most favorable for cooling. The arrows indicate the shift between the two beam centers.

Cooling with an offset provides sufficient rate of the emittances decrease to permit efficient transfers from the Accumulator, while the lifetime does not deteriorate as much as it would with on-axis cooling. When the stored beam needs to be prepared for extraction, the main goal for cooling is to reach as small emittances as possible without inducing an instability [55]. Typically, it requires the electron beam to be brought closer to the antiproton beam central orbit. Over the relatively short time it takes to extract the antiprotons, the life time is not a critical issue.

## *10.3.    Impact on the Tevatron complex performance*

Electron cooling was one of many improvements carried out during Run II. The most direct effect on the Tevatron luminosity is the increase of the antiproton beam phase density during the preparation of the stacks for extraction to the Main Injector (followed by injection in the Tevatron). By applying strong electron cooling right before extraction the transverse emittances decreased by a factor of ~1.5 with respect to what it was possible to do with stochastic cooling alone, and for a much larger number of antiprotons (~$400 \times 10^{10}$ vs. ~$150 \times 10^{10}$ antiprotons, typically). Ref. [57] estimates the corresponding luminosity improvement to be ~25% (p. 16).

Also, the extraction efficiency (from the Recycler to the Main Injector) was noticeably increased to ~98% on average and more than 99% for small stacks thanks to the smaller longitudinal emittance of the stored beam. In addition, the smaller longitudinal emittance of the extracted bunches led to higher coalescing efficiencies in the Main Injector. Because these changes were implemented in parallel with other improvements in the complex, the actual impact on luminosity is difficult to evaluate quantitatively but is likely ~10%.

Besides to these direct contributions, two indirect effects can be identified as being critical for improving the luminosity potential even further.

First, electron cooling essentially removed the limit on the amount of antiprotons that was possible to store and prepare in the Recycler for the following transfers to the Main Injector and the Tevatron collider. The record high number of antiprotons stored in the Recycler was $606 \cdot 10^{10}$ (March 21, 2011), without any clear indications of strong limiting factors. At the same time the typical Recycler stack size before extraction was $\sim 380 \cdot 10^{10}$ and was dictated by optimization in other machines.

Second, fast longitudinal cooling of the bunches coming from the Antiproton Accumulator allowed much more frequent transfers. As a result, the average size of the Accumulator stack decreased and in turn, led to an increase of the antiproton production efficiency.

Note that electron cooling also brought more flexibility to beam manipulations in the Recycler (in the longitudinal direction) and to the way the accelerator complex could be and was operated.

As an indication of the importance of electron cooling one can consider the drop of the integrated luminosity rate without it. In several instances toward the end of the Run II when the cooler was not operational for 2-3 days, the rate of integrating luminosity would decrease by a factor of ~2.5.

## 11. Final remarks

The Recycler Electron cooler was the first and, during its operation, the only relativistic ($\gamma$>2) electron cooler. It significantly contributed to the success of Run II, resulting in an immediate increase of the integrated luminosity by ~25% and even more by removing a bottleneck in the antiproton production.

In addition, technical solutions found during the project and better understanding of the peculiarities of the cooling process should help to build future coolers.

## 12. Acknowledgement

While this paper was compiled by two authors, a much larger group has conceived, designed, built, commissioned, and maintained the Recycler Electron cooler as well as participated in measurements and formulating results. Major contributions by scientists and engineers are reflected by the authorship of the referenced papers; however, support from many others at Fermilab and from the international scientific community had a great cumulative effect. Also, the success of the Fermilab electron cooler would have not been possible without the dedicated work of the technical personnel of the Accelerator Division.

# 13. Bibliography

[1] G. Budker, *Sov. Atomic Energy,* vol. 22, p. 346, 1967.

[2] G. I. Budker and e. al, *IEEE Trans. Nucl. Sci.,* Vols. NS-22, p. 2093, 1975.

[3] I. N. Meshkov, *Phys. Part. Nucl.,* vol. 25, no. 6, p. 631, 1994.

[4] "Fermilab Recycler Ring Technical Design Report FERMILAB-TM-1991," 1996. [Online]. Available: http://lss.fnal.gov/archive/test-tm/1000/fermilab-tm-1991.pdf.

[5] S. Nagaitsev et al, *Phys. Rev. Lett.,* vol. 96, p. 044801, 2006.

[6] Y. Derbenev and A. N. Skrinsky, *Particle Accelerators,* vol. 8, p. 1, 1977.

[7] I. N. Meshkov et al, "Physics guide of BETACOOL code v.1.1," BNL note C-A/AP#262 p.17.

[8] A. Shemyakin and L. R. Prost, "Electron Cooling Rates in FNAL's Recycler Ring," in *Proc. of APAC'07*, Indore, India, 2007, TUPMA094.

[9] J. A. MacLachlan, A. Burov, A. C. Crawford, T. Kroc, S. Nagaitsev, C. Schmidt, A. Sharapa, A. Shemyakin and A. Warner, "Prospectus for an electron cooling system for the Recycler," FERMILAB-TM-2061, 1998.

[10] M. Veis et.al, in *Proc. of EPAC'88*, Rome, Italy, 1988, p.1361..

[11] J. Adney et al, in *Proc. of PAC'89*, Chicago, USA, 1989, p.348.

[12] G.P.Jackson, "Modified Betatron Approach to Electron Cooling," in *MEEC'97*, Novosibirsk, 1997.

[13] I. Ben-Zvi, in *Proc. of EPAC'06*, Edinburgh, UK, 2006, TUZBPA01.

[14] J. Dietrich et al, in *Proc. Of COOL'11*, Alushta, Ukraine, 2011, MOIO05.

[15] D. R. Anderson et al, in *Proc. of EPAC'92*, 1992, p.836-838.

[16] N. Dikansky et al, "Electron beam focusing system," FERMILAB-TM-1998, 1996.

[17] A. Burov, S. Nagaitsev, A. Shemyakin and Y. Derbenev, "Optical principles of beam transport for relativistic electron cooling," *Phys. Rev. Special Topics: Accelerators and Beams,* vol. 3, p. 094002, 2000.

[18] Pelletrons are manufactured by the National Electrostatics Corporation, www.pelletron.com.

[19] A. Sharapa, A. Shemyakin and S. Nagaitsev, *Nucl. Instr. and Meth.A,* vol. 417, p. 177, 1998.

[20] A. Burov, T. Kroc, V. Lebedev, S. Nagaitsev, A. Shemyakin, A. Warner and S. Seletskiy, "Optics of the Fermilab electron cooler," in *Proc. of APAC'04*, Gyeongju, Korea, 2004, pp. 647-649.

[21] V. Lebedev, "OptiM - Computer code for linear and non-linear optics calculations," [Online]. Available: http://www-bdnew.fnal.gov/pbar/organizationalchart/lebedev/OptiM/optim.htm.

[22] P. W. Joireman et al, "BPM System for Electron Cooling in the Fermilab Recycler Ring," in *Proc. of Beam Instrumentation Workshop 2004, AIP Conf. Proc. 732*, 2004, 319.

[23] The program for data recording was written by V. Lebedev and V. Nagaslaev.

[24] A. V. Ivanov and M. A. Tiunov, "ULTRASAM - 2D Code for Simulation of Electron Guns with Ultra High Precision," in *Proc. of EPAC-2002*, Paris, 2002.

[25] M. A. Tiunov, "BEAM − 2D-code package for simulation of high perveance beam dynamics in long systems," in *Proc. of SPACE CHARGE EFFECTS IN FORMATION OF INTENSE LOW ENERGY BEAMS*, Dubna, Russia, 1999.

[26] A. Warner et al, "OTR Measurements and Modeling of the Electron Beam Optics at the E-cooling Facility," in *AIP Conf.Proc. 821*, 2006, pp. 380-385.

[27] T. Kroc et al, "Electron beam size measurements in a cooling solenoid," in *Proc. of PAC'05*, Knoxville, USA, 2005, p.3801.

[28] A. Burov et al, "Optics of electron beam in the Recycler," in *Proc. of COOL'05*, Galena, USA, 2005, p. 139.

[29] A. Shemyakin, "Electron Beam Collector with a Transverse Magnetic Field," in *Proc. of EPAC'00*, Vienna, 2000.

[30] L. R. Prost and A. Shemyakin, "Efficiency of the Fermilab Electron cooler's collector," in *Proc. of PAC'05*, Knoxville, USA, 2005, p.2387.

[31] A. Burov et al, "IBS in a CAM-dominated electron beam," in *Proc. of COOL'05*, Galena, USA, 2005, p. 159, also available as FERMILAB-CONF-05-462-AD.

[32] A. Shemyakin et al, "Attainment of an MeV-range, DC electron beam for the Fermilab cooler," in *Proc. of COOL'03*, Japan, 2003.

[33] L. R. Prost and A. Shemyakin, "Full discharges in the Fermilab Electron cooler," in *Proc. of COOL'05*, Galena, USA, 2005, p. 391.

[34] A. Warner et al, "The design and implementation of the machine protection system for the Fermilab electron cooling facility," in *Proc. of DIPAC-09*, 2009.

[35] L. R. Prost et al, "Electron cooling status and characterization at Fermilab's Recycler," in *Proc. of COOL'07*, Bad Kreuznach, Germany, 2007, MOA2I04.

[36] A. Shemyakin et al, "Optimization of Electron Cooling in the Recycler," in *Proc. of PAC'09*, Vancouver, Canada, 2009, TU6PFP076.

[37] A. Burov, "Optical Linearity of Ecool Line, Report at departmental meeting," available at http://beamdocs.fnal.gov/AD-public/DocDB/ShowDocument?docid=2966, 2007.

[38] A. Shemyakinet al, "Effect of Secondary Ions on the Electron Beam Optics in the Recycler Electron Cooler," in *Proc. of IPAC'10*, Kyoto, Japan, 2010, MOPD075.

[39] *The measurements wer made by P. Joireman, and the analysis was performed by A. Burov.*

[40] A. Burov, "ECool BPM Noise Analysis, Report at departmental meeting," available at http://beamdocs.fnal.gov/AD-public/DocDB/ShowDocument?docid=3099, 2007.

[41] G. Kazakevich et.al, "Recycler Electron Cooling project: Mechanical vibrations in the Pelletron and their effect on the beam," FERMILAB-TM-2319-AD, 2005.

[42] A. Crawford et.al, "Field measurements in the Fermilab electron cooling solenoid prototype," Fermilab preprint TM-2224, 2003.

[43] V. Tupikov et.al, "Magnetic Field Measurement and Compensation in Recycler Electron Cooler," in *Proc. of COOL'05*, Galena, USA, 2005, p. 375.

[44] A. Shemyakin et al, "Stability of electron energy in the Fermilab Electron cooler," Fermilab preprint CONF-08-425-AD, 2008.

[45] S. M. Seletskiy and A. Shemyakin, "Beam-based calibration of the electron energy in the Fermilab electron cooler," in *Proc. of PAC'05*, Knoxville, USA, 2005, p. 3638.

[46] S. Nagaitsev et.al, "Antiproton cooling in the Fermilab Recycler," in *Proc. of COOL'05*, Galena, USA, 2005, p.3638.

[47] A. Khilkevich, L. R. Prost and A. Shemyakin, "Effect of Transverse Electron Velocities on the Longitudinal Cooling Force in the Fermilab Electron Cooler," in *Proc. of PAC'11*, New York, USA, 2011, WEP228.

[48] L.R.Prost and A. Shemyakin, "Electron cooling rates characterization at Fermilab's Recycler," in *Proc. of PAC'07*, Albuquerque, USA, June 25- 29, 2007, TUPAS030.

[49] A. Shemyakin et.al., "Electron cooling in the Recycler cooler," in *COOL'07*, Bad Kreuznach, Germany, September 10-14, 2007.

[50] D. Reistad and L. Hermansson, *NIM A 441,* p. 140, 2000.

[51] A. Burov and S. Nagaitsev, "Envelope instability as a source of diffusion," *NIM* , vol. A441, pp. 18-22, 2000.

[52] A. Shemyakin and A. Valishev, "Interaction between protons and electrons in the Recycler," Report at the departmental meeting, http://beamdocs.fnal.gov/AD-public/DocDB/ShowDocument?docid=3554, 2010.

[53] A. Shemyakin et al, "Attainment of high-quality electron beam for Fermilab 4.3-MV cooler," in *Proc. of COOL'05*, Galena, IL USA, 2005, p. 280.

[54] A. Bubley et .al, "The electron gun with variable beam profile for optimization of electron cooling," in *Proc. of EPAC'02*, Paris, France, 2002, WEPRI049.

[55] L. Prost et al, "Transverse Instability of the Antiproton Beam In the Recycler Ring," in *Proc. of PAC'11*, New York, USA, 2011, WEP114.

[56] S. Holmes, R. S. Moore and V. Shiltsev, "Overview of the Tevatron Collider Complex: Goals, Operations and Performance," 2011. [Online]. Available: http://arxiv.org/ftp/arxiv/papers/1106/1106.0909.pdf.